\documentclass[a4paper,fleqn]{cas-dc}

\usepackage[numbers]{natbib}
\usepackage{lipsum}
\usepackage{siunitx}
\usepackage{subcaption}
\usepackage{makecell}
\usepackage[switch]{lineno}

\definecolor{LightYellow}{RGB}{255,255,128}
\usepackage{soul}
\setul{-1.5ex}{2ex}
\setulcolor{LightYellow}

\usepackage{courier}
\usepackage{silence}
\def\tsc#1{\csdef{#1}{\textsc{\lowercase{#1}}\xspace}}
\tsc{WGM}
\tsc{QE}
\tsc{EP}
\tsc{PMS}
\tsc{BEC}
\tsc{DE}

\begin{document}
\let\WriteBookmarks\relax
\def\floatpagepagefraction{1}
\def\textpagefraction{.001}

\shorttitle{The role of data partitioning on the performance of EEG-based deep learning models in cross-subject analysis}  

\shortauthors{F. Del Pup et~al.}  

\title [mode = title]{The role of data partitioning on the performance of EEG-based deep learning models in supervised cross-subject analysis: a preliminary study} 

\tnotemark[1] 

\tnotetext[1]{
This document is the results of the research project funded
in part by the European Union’s Horizon Europe research and innovation programme under Grant agreement no 101137074 - HEREDITARY,
in part by the STARS@UNIPD funding program of the University of Padua, Italy, project: MEDMAX.
}


\author[1,2,3]{Federico {Del Pup}}[
    type=author,
    auid=000,
    bioid=1,
    orcid=0009-0004-0698-962X
]
\cormark[1]


\ead{federico.delpup@studenti.unipd.it}


\credit{
    Conceptualization,
    Methodology,
    Software,
    Visualization,
    Writing – original draft
}

\author[2,3]{Andrea {Zanola}}[
    type=author,
    auid=001,
    bioid=2,
    orcid=0000-0001-6973-8634
]
\ead{andrea.zanola@phd.unipd.it}
\credit{
    Methodology,
    Writing - review and editing
}

\author[2,3]{{Louis Fabrice} Tshimanga}[
    type=author,
    auid=002,
    bioid=3,
    orcid=0009-0002-1240-4830
]
\ead{louisfabrice.tshimanga@unipd.it}
\credit{
    Methodology,
    Writing - review and editing
}

\author[1,3]{Alessandra {Bertoldo}}[
    type=author,
    auid=003,
    bioid=4,
    orcid=0000-0002-6262-6354
]
\ead{alessandra.bertoldo@unipd.it}
\credit{Writing - review and editing}

\author[3,4,5]{Livio {Finos}}[
    type=author,
    auid=004,
    bioid=5,
    orcid=0000-0003-3181-8078
]
\ead{livio.finos@unipd.it}
\credit{
    Methodology,
    Writing - review and editing
}

\author[2,3,6]{Manfredo {Atzori}}[
    type=author,
    auid=005,
    bioid=6,
    orcid=0000-0001-5397-2063
]
\ead{manfredo.atzori@unipd.it}
\credit{Supervision, Funding acquisition, Project administration, Writing - review and editing}

\affiliation[1]{
    organization={Department of Information Engineering, University of Padua}, 
    city={Padua},
    postcode={35131}, 
    country={Italy}
}

\affiliation[2]{
    organization={Department of Neuroscience, University of Padua}, 
    city={Padua},
    postcode={35121}, 
    country={Italy}
}

\affiliation[3]{
    organization={Padova Neuroscience Center, University of Padua}, 
    city={Padua},
    postcode={35129}, 
    country={Italy}
}

\affiliation[4]{
    organization={Department of Statistical Sciences, University of Padua}, 
    city={Padua},
    postcode={35121}, 
    country={Italy}
}

\affiliation[5]{
    organization={Department of Developmental and Social Psychology, University of Padua}, 
    city={Padua},
    postcode={35131}, 
    country={Italy}
}

\affiliation[6]{
    organization={Information Systems Institute, University of Applied Sciences Western Switzerland (HES-SO Valais)}, 
    city={Sierre},
    postcode={3960}, 
    country={Switzerland}
}

\cortext[1]{Corresponding author}



\begin{abstract}
Deep learning is significantly advancing the analysis of electroencephalography (EEG) data by effectively discovering highly nonlinear patterns within the signals.

Data partitioning and cross-validation are crucial for assessing model performance and ensuring study comparability, as they can produce varied results and data leakage due to specific signal properties (e.g., biometric).
Such variability in model evaluation leads to incomparable studies and, increasingly, overestimated performance claims, which are detrimental to the field.  
Nevertheless, no comprehensive guidelines for proper data partitioning and cross-validation exist in the domain, nor is there a quantitative evaluation of the impact of different approaches on model accuracy, reliability, and generalizability.

To assist researchers in identifying optimal experimental strategies, this paper thoroughly investigates the role of data partitioning and cross-validation in evaluating EEG deep learning models.

Five cross-validation settings are compared across three supervised cross-subject classification tasks (brain-computer interfaces, Parkinson’s, and Alzheimer’s disease classification) and four established architectures of increasing complexity (ShallowConvNet, EEGNet, DeepConvNet, and Temporal-based ResNet).

The comparison of over 100,000 trained models underscores, first, the importance of using subject-based cross-validation strategies for evaluating EEG deep learning architectures, except when within-subject analyses are acceptable (e.g., BCI). 
Second, it highlights the greater reliability of nested approaches (e.g., N-LNSO) compared to non-nested counterparts, which are prone to data leakage and favor larger models overfitting to validation data.

In conclusion, this work provides EEG deep learning researchers with an analysis of data partitioning and cross-validation and offers guidelines to avoid data leakage, currently undermining the domain with potentially overestimated performance claims.
\end{abstract}



\maketitle

\section{Introduction}
\label{sec: intro}

Over the past decade, Deep Learning (DL) has dramatically transformed the analysis of electroencephalographic (EEG) data, achieving state-of-the-art performance across a variety of applications, including Brain-Computer Interfaces (BCI) \cite{BCIrew}, emotion recognition \cite{emorew}, sleep staging \cite{sleeprew}, and the characterization of neurological disorders such as Parkinson's \cite{ParkinsonRew}, Alzheimer's \cite{Alzrew}, and epilepsy \cite{EpilepsyRew}.
This success stems from the ability of deep neural networks to extract complex nonlinear patterns that reflect the brain neural activity.
However, the potential of deep learning is significantly hindered by its sensitivity to minimal variations in the experimental setting.
For instance, the use of extensive preprocessing pipelines---though supported by literature due to the low signal-to-noise ratio of EEG signals---can lead to significant performance drops of over 10\% compared to models trained with minimally filtered data \cite{DelPup2024b}.
Similarly, the methodology used to partition data into training, validation, and test sets can strongly influence the final results, leading to unrealistic estimates of performance metrics and a lack of comparability across studies.

In EEG analysis, the limited availability of datasets with more than a few dozen or, at most, a hundred subjects, constrains the creation of sufficiently large and representative training, validation, and test sets. 
Consequently, it has become best practice to assess model accuracy using Cross-Validation (CV) procedures, which yield more reliable performance estimates, particularly when working with small datasets.
Over the years, researchers have adopted various cross-validation strategies that operate both at the sample and the subject level \cite{Xvaltypes}.
Sample-level CVs allow segments of the same EEG recording to be assigned in both the training and the validation set, while subject-level CVs ensure that data from individual subjects are not mixed between sets.

As noted in \cite{SubjAwSSL, kamrudloso}, biosignals like EEG are characterized by strong subject-specific characteristics that can be learned during training and exploited by the model during inference.
This can lead to overfitting and unrealistic performance estimates on unseen subjects.
However, sample-based methods remain common in deep learning-based EEG data analysis.
According to Brookshire et al. {\cite{sample_leakage_2}}, who reviewed a total of 63 deep learning studies on translational EEG, only a small percentage (27.0\%) unambiguously avoided this type of data leakage.
Consequently, there has been a growing emphasis on subject-based cross-validation methods, such as the Leave-N-Subjects-Out (LNSO) or the Leave-One-Subject-Out (LOSO), which aim to develop generalizable models that are less sensitive to the high inter-subject variability.

Quantitative analyses have been conducted to evaluate differences between traditional sample-based and subject-based cross-validation methods, underscoring the importance of using the latter in EEG deep learning studies \cite{LOSO, LOSO2}.
Nonetheless, even standard subject-based CV methods do not align well with DL paradigms.
To avoid overfitting on the training data, EEG-DL models necessitate a separate independent set for early stopping purposes.
This is crucial for estimating the generalization error during training and halting the process if improvement does not occur over a specified number of epochs.
Unfortunately, this process is frequently conducted on the validation set, which is subsequently used to evaluate the model, leading to data leakage and reinforcing overfitting tendencies \cite{noval1, noval2, noval3}.
To address this issue, \cite{nloso_emo} proposed a nested subject-based CV approach, characterized by separate validation and test sets.
This methodology, formalized in \cite{DelPup2024b} as Nested-Leave-N-Subjects-Out (N-LNSO), allowed to move to a train-validation-test framework, which is more suited for EEG-DL applications, albeit with increased computational costs.

The variety of possible cross validation variants leads to inconsistencies in results, complicating comparisons across studies and causing uncertainty on the selection of an optimal approach.
Despite this, a comprehensive evaluation of the precise impact of this important aspect is still missing.

\textbf{Contributions}:
This work aims at thoroughly investigating the role of data partition on the performance assessment of EEG-DL models through cross-validation analysis.
The investigation involves a comparison of five distinct cross-validation settings from three main categories, as described in subsection {\ref{subsec: CV}}.
These categories encompass both sample-based and subject-based strategies.
The comparison is conducted across three supervised cross-subject classification tasks (brain-computer interface, Parkinson’s disease, and Alzheimer’s disease classification) using four established EEG-DL architectures of increasing complexity (ShallowConvNet, EEGNet, DeepConvNet, and Temporal-based 1D ResNet).
Initially, traditional sample-based and subject-based CVs are compared to determine the extent to which model inference is affected by subject-specific characteristics, potentially introducing bias.
Subsequently, traditional and nested subject-based CVs are compared to evaluate the consequences of lacking independent validation and test sets.
Finally, variations in accuracies derived from nested and non-nested subject-based cross-validation strategies are analyzed in relation to model complexity, assessing whether certain approaches potentially favor more complex models.

\textbf{Paper structure}: The outline of this paper is as follows.
Section \ref{sec: methods} describes in detail the experimental setting.
Section \ref{sec: results} presents the results, which are discussed with their limitations in section \ref{sec: discussion}.
Finally, a conclusion is drawn in section \ref{sec: conclusion}.

\section{Methods}
\label{sec: methods}

This section describes in detail important methodological aspects.
It follows the checklist provided in \cite{DelPup2024a}, which lists all the features a research work should report to improve result reproducibility without affecting readability.
These include: dataset selection, data preprocessing, models' architecture, data partitioning and cross-validation schemes, training hyperparameters, and model evaluation.
Additional details can be found within the supplementary materials and the openly available source code repository.

\subsection{Dataset selection}
\label{subsec: dataset}
\label{data-sel}
The analysis was conducted on four different open-source datasets. 
Each of them was selected from OpenNeuro\footnote{[Online] Available: \href{https://openneuro.org/}{https://openneuro.org/}} \cite{OpenNeuro}, an established open platform used by neuroscientists to share neuroimaging data associated to published research studies, with a digital object identifier.
This platform was selected because it specifically requires data to be organized in BIDS format, a standardized data structure format for neuroimaging data \cite{EEGBIDS}.
Such feature facilitates the exploitation of automated preprocessing and harmonization tools such as \cite{BIDSAlign}, enhancing study reproducibility.  
Furthermore, all selected datasets include raw EEG data recorded at least with the minimum number of channels suggested in the American Clinical Neurophysiology Society (ACNS) guidelines \cite{eeguide}, i.e., 19 plus 2 for contralateral referencing. 

Selected data were used to construct three different and widely investigated classification tasks, namely:
\WarningsOff
\begin{enumerate}[\textbullet]
    \item \textit{Parkinson}: a binary pathology classification task. It aims to distinguish between Parkinson's off-medication and healthy subjects. EEG has been investigated as a non-invasive, low-cost technique to be used in Parkinson's clinical diagnosis, with deep learning applications producing promising results \cite{ParkinsonRew}.
    \item \textit{Alzheimer}: a three-classes pathology classification task. It aims to distinguish between control, Alzheimer's, and Frontotemporal Dementia subjects. Similar to Parkinson's applications, deep learning has been recently investigated as a promising tools for analyzing the nonlinear dynamics of EEGs that characterize neurological disorders \cite{Alzrew}.
    \item \textit{BCI}: a non-clinical four-classes classification task. It aims to distinguish between left/right imagined/executed fist's movements. BCI is a widely studied EEG application that found potential usages both in industry (e.g., operative security in critical conditions) and in medicine (e.g., stroke rehabilitation) \cite{bci_40, bci_med}.
\end{enumerate}
\WarningsOn
A preliminary analysis with two additional tasks is provided in the Supplementary Material (Section A.7).
Further acquisition details for each dataset are summarized in \autoref{tab: datasets}.
The next sub-paragraphs concisely describe them.

\renewcommand{\arraystretch}{1.2} 
\begin{table*}[!th]
\centering
\caption{Dataset description.}
\begin{tabular}{ccccccccc}
\toprule
\makecell{dataset\\ID} & \makecell{original\\reference} & \makecell{number\\of\\
channels} & \makecell{original\\sampling\\rate [Hz]} & \makecell{number\\of\\subjects} & \makecell{number\\of\\windows} & \makecell{number\\of\\classes} & \makecell{task\\description} & \makecell{task\\acronym} \\ \midrule
\multicolumn{1}{c}{ds004362} & \multicolumn{1}{c}{A1-A2} & \multicolumn{1}{c}{64} & \multicolumn{1}{c}{160} & \multicolumn{1}{c}{109$^{*}$} & \multicolumn{1}{c}{9495} & \multicolumn{1}{c}{4} & \multicolumn{1}{c}{\makecell{Motor/Imagery\\left vs right fist}} & BCI \\
\multicolumn{1}{c}{ds002778} & \multicolumn{1}{c}{CMS-DRL} & \multicolumn{1}{c}{41} & \multicolumn{1}{c}{512} & \multicolumn{1}{c}{31} & \multicolumn{1}{c}{\multirow{2}{*}{8608}} & \multicolumn{1}{c}{\multirow{2}{*}{2}} &  \multicolumn{1}{c}{\multirow{2}{*}{\makecell{Control vs Parkinson's\\off-medication}}} & \multirow{2}{*}{Parkinson} \\ 
\multicolumn{1}{c}{ds003490} & \multicolumn{1}{c}{CPZ} & \multicolumn{1}{c}{64} & \multicolumn{1}{c}{500} & \multicolumn{1}{c}{50} & \multicolumn{1}{c}{} & \multicolumn{1}{c}{} &  \\
\multicolumn{1}{c}{ds004504} & \multicolumn{1}{c}{A1-A2} & \multicolumn{1}{c}{19} & \multicolumn{1}{c}{500} & \multicolumn{1}{c}{88} & \multicolumn{1}{c}{17252} & \multicolumn{1}{c}{3} & \multicolumn{1}{c}{\makecell{Control vs Alzheimer\\vs FT-dementia}} & Alzheimer \\ 
\bottomrule
\multicolumn{9}{l}{\text{*three subjects were excluded from the analysis due to inconsistent sampling rate and trial length}}
\end{tabular}
\label{tab: datasets}
\end{table*}
\renewcommand{\arraystretch}{1} 

\subsubsection{Parkinson's 1 - ds002778}
This dataset \cite{ds002778} collects resting-state eyes open EEG recordings (duration $195.7 \pm 18.8$ seconds) from 15 Parkinson's patients and 16 age matched healthy controls (respectively $63.3 \pm 8.2$ years, $63.5 \pm 9.7$ years).
Healthy subjects have only one session, while Parkinson's have two: the first contains records from Parkinson's patients who discontinued medication for at least 12 hours before the session (\textit{ses-off}), while the second include records of the same patients under medication (\textit{ses-on}).
For the following analysis, only the off-medication session was included.
In addition, since the number of subjects is too low to perform proper data partition strategies, this dataset is used together with the Parkinson's 2 dataset (ds003490), described below.

\subsubsection{Parkinson's 2 - ds003490}
This dataset \cite{ds003490} contains resting-state eyes open/closed and auditory oddball EEG recordings (duration $595.9 \pm 74.0$ seconds) from 25 Parkinson's patients and 25 age matched healthy controls (respectively $69.7 \pm 8.7$ years, $69.3\pm 9.6$ years).
Similarly to the previous dataset, healthy subjects have only one session, while Parkinson's patients have off-medication (at least 15 hours) and on-medication recordings.

\subsubsection{Alzheimer's - ds004504}
\label{ds004504}
This dataset \cite{ds004504} collects resting-state eyes closed EEG recordings (duration $802.2 \pm 140.3$ seconds) from 23 subjects diagnosed with frontotemporal dementia, 36 Alzheimer's patients, and 29 age matched healthy controls (respectively $63.7\pm8.2$ years, $66.4 \pm 7.9$ years, and $67.9\pm5.4$ years).
All the subjects have only one session, which was used in this work.

\subsubsection{BCI - ds004362}
This dataset contains resting-state, motor movement, and motor imagery records of 109 adult healthy subjects (age $39 \pm 11$ years).
All subjects have multiple records acquired during a single session (duration $114.5 \pm 22.0$ seconds).
In particular, there are two resting-state eyes open/closed EEG acquisition and three repetitions of four different trial-based tasks, namely:
\begin{enumerate}[{Task} 1.]
    \item Open and close left or right fist.
    \item Imagine opening and closing left or right fist.
    \item Open and close both fists or feet.
    \item Imagine opening and closing both fists or feet.
\end{enumerate} 
Trials from tasks 1 and 2 were considered to perform the `BCI' task.
Furthermore, subjects with ID 88, 92 and 100, have been excluded due to inconsistent sampling rate and trial length, as already done in other works \cite{eegnetfam, mmi_rm2}.

\subsection{Data preprocessing}
\label{subsec: prepro}

All the selected recordings were preprocessed with task-specific pipelines. 
Each of them reflects the best possible configuration according to a recent analysis on the role of preprocessing in EEG-DL applications \cite{DelPup2024b}.
Pipelines are composed by common steps adopted by both clinicians \cite{happe} and deep learning practitioners \cite{Yannick}, and are summarized in \autoref{tab: pipelines}.
Preprocessing was done with \textit{BIDSAlign}\footnote{https://github.com/MedMaxLab/BIDSAlign} \cite{BIDSAlign} (v1.0.0), with the sole exception for the final standardization, downsampling, and windows extraction steps, which were performed online within the Python environment as part of the data loading operations during model training.
A detailed description of each step and relative parameters is provided below, sorted by their order of execution.

\renewcommand{\arraystretch}{1.2} 
\begin{table}[!th]
    \centering
    \caption{Preprocessing pipelines for each task.}
    \begin{tabular}{cccc}
        \toprule
        preprocessing step & BCI & Parkinson & Alzheimer \\
        \midrule
        \multicolumn{1}{c}{\makecell{non-EEG channels\\removal}} & \checkmark & \checkmark & \checkmark \\
        \multicolumn{1}{c}{time segments removal} &  & \checkmark & \checkmark \\
        \multicolumn{1}{c}{DC component removal} & \checkmark & \checkmark & \checkmark \\
        \multicolumn{1}{c}{resampling} &  & \checkmark & \checkmark \\
        \multicolumn{1}{c}{filtering} & \checkmark & \checkmark & \checkmark \\
        \multicolumn{1}{c}{\makecell{automatic independent\\component rejection}} &  & \checkmark & \checkmark \\
        \multicolumn{1}{c}{bad-channel removal} &  &  & \checkmark \\
        \multicolumn{1}{c}{\makecell{bad-time windows\\correction with ASR}} &  &  & \checkmark \\
        \multicolumn{1}{c}{\makecell{spherical interpolation\\of removed bad-channels}} &  &  & \checkmark \\ 
        \multicolumn{1}{c}{re-reference} & \checkmark & \checkmark & \checkmark \\
        \multicolumn{1}{c}{template alignment} & \checkmark & \checkmark &  \\
        \multicolumn{1}{c}{trials extraction} & \checkmark &  &  \\
        \multicolumn{1}{c}{standardization} & \checkmark & \checkmark & \checkmark  \\
        \multicolumn{1}{c}{downsampling} & \checkmark & \checkmark & \checkmark  \\
        \multicolumn{1}{c}{windows extraction} &  & \checkmark & \checkmark  \\
        \bottomrule
    \end{tabular}
    \label{tab: pipelines}
\end{table}
\renewcommand{\arraystretch}{1} 

\WarningsOff
\begin{enumerate}[$\bullet$]
    \item \textit{non-EEG channels removal}: other biosignals stored together with the EEG and usually included as extra channels have been removed.
    \item \textit{time segments removal}: the first and last 8 seconds of each recording have been removed to exclude potential divergences in the signal caused, for example, by the instrumentation. This step makes more accurate the subsequent artifact handling operations, which can potentially remove important parts of the brain activity when the input signal is very noisy.
    \item \textit{DC component removal}: The channel-wise direct current (DC) voltage was subtracted from each EEG. More formally, denoting the original EEG signal by a matrix $\bold{X}\in \mathbb{R}^{C\times T}$, where C represents the number of channels and T the sequence length, the new DC-subtracted signal ${\bold{X}}^\prime\in~\mathbb{R}^{C\times{T}}$ can be defined as:
    \begin{equation}
    \label{eq:DCrem}
        {\bold{X}}^\prime = \bold{X} - \bar{\mathbf{x}}_{c}\textit{diag}(\bold{I}_T)
    \end{equation}
    with $\bar{\mathbf{x}}_{c} \in \mathbb{R}^{C\times 1}$ the vector of means of the channels. 
    \item \textit{resampling}: EEG records with a sampling rate higher than 250Hz were resampled to 250Hz.
    \item \textit{filtering}: EEG signals were filtered with a pass-band Hamming windowed sinc FIR filter between 1Hz and 45Hz.
    \item \textit{automatic independent component rejection}: Independent Components (IC) were extracted using the `\textit{runica}' algorithm {\cite{infomaxica}}, with no limit on their number; then, components were automatically rejected using IClabel {\cite{ICLabel}}, a deep learning classifier commonly used in state-of-the-art EEG preprocessing pipelines.
    ICLabel calculates the probability of each IC belonging to one of seven categories: brain, muscle, eye, heart, line noise, channel noise, or other.
    If the calculated probability falls within a predefined range set by the user, the IC is rejected.
    For this study, the rejection thresholds were set as follows: [90\%, 100\%] for noisy classes and [0\%, 10\%] for the brain category.
    These values were selected to ensure the rejection of both predefined and other artifacts without excessive removal of brain activity.
    The number of rejected components may vary depending on the quality of the EEG signal.
    \item \textit{bad-channel removal}: flat or extremely noisy channels are removed using the `\textit{clean\_rawdata}' plugin (a BIDSAlign dependency) with default parameters.
    \item \textit{bad-time windows correction}: the Artifact Subspace Reconstruction (ASR) algorithm \cite{asr} with default parameters was used to detect and correct bad-time windows, i.e. particularly noisy portions of the signal.
    \item \textit{spherical interpolation}: previously removed bad channels are interpolated with the spherical method \cite{Spherical}.
    \item \textit{re-reference}: EEG recordings are re-referenced to the common average.
    \item \textit{template alignment}: EEG signals collected in the Parkinson's 1 dataset have a lower channel count compared to the Parkinson's 2. For this reason, only channels in common were selected. 
    In addition, data for the `\textit{BCI}' task were aligned to the International Federation of Clinical Neurophysiology (IFCN) 10-10 standard containing 61 channel, thus removing 3 channels.
    \item \textit{trials extraction}: BCI trials were extracted using the event files provided by original authors. Whenever possible, if the trial was few samples shorter, a piece of the next adjacent resting-state part of the EEG signal (rest portion between trials) was included to make the trial long 4.1 seconds, which is the intended length.
    \item  \textit{downsampling}: During the import of preprocessed files within the Python environment, data were further downsampled to 125 Hz to improve computation and reduce the GPU's memory occupation. BCI trials, which have an original sampling rate that is not multiple of 125Hz, have been resampled with the windowed sinc interpolation method, Von Han window, provided by the `\textit{torchaudio}' library \cite{torchaudio}.
    \item \textit{standardization}: a final z-score operator was applied along the EEG channel dimension. This step transforms each EEG channel into a signal with $\mu\text{=0}$ and $\sigma\text{=1}$.
    \item \textit{windows extraction}: EEG data for the Parkinson's and Alzheimer's tasks were partitioned into non-overlapping 4s windows. This step was not applied to BCI data due to the trial extraction step.
\end{enumerate}
\WarningsOn

\subsection{Cross-validation strategies}
\label{subsec: CV}
\begin{figure*}[!t]
    \centering
    \includegraphics[trim={1,5cm 0cm 0cm 0cm}, width=\textwidth]{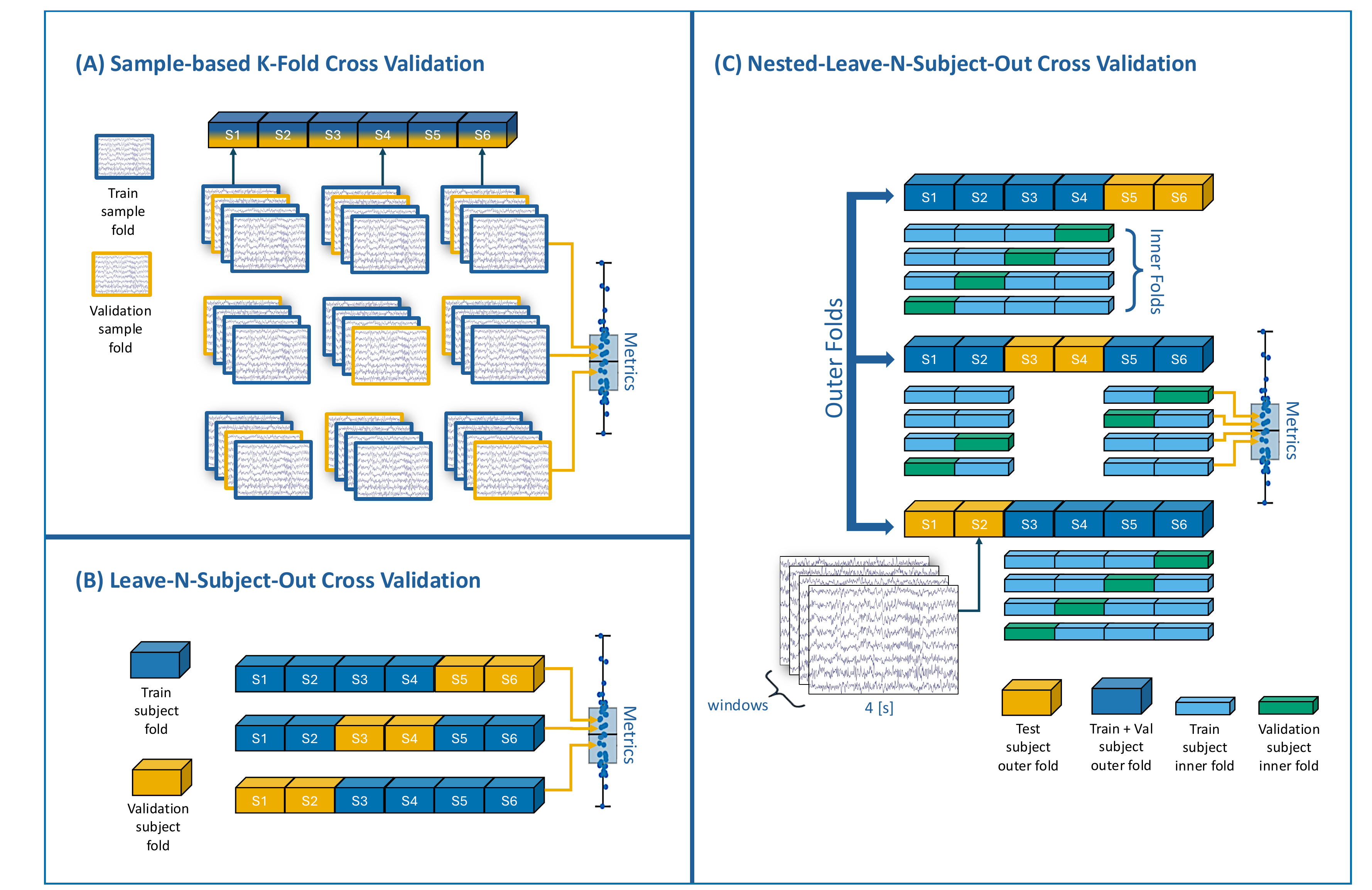}
    \caption{
    Schematic representation of the three categories of cross-validation investigated in this work, inspired by \cite{DelPup2024b}.
    (a) Sample-based K-Fold (K-Fold) randomly assigns windows of the same EEG both in the training and in the validation set, ignoring the subject;
    (b) Leave-N-Subjects-Out (LNSO or LOSO if N=1) randomly assigns windows of the same subject either in the training or in the validation set (not in both);
    (c) Nested-Leave-N-Subjects-Out (N-LNSO or N-LOSO) introduces a nested level for the creation of multiple train-validation splits for each test split.
    Models are trained on the training set, and evaluated on test set, if available; otherwise, the validation set is used.
    This procedure generates an ensemble of performances, which is used for the subsequent cross-validation comparisons.
    The training is controlled using early stopping criteria on the validation set.
    Note that for (a) and (b), the usage of the validation set to both monitor the training and evaluate the model introduces data leakage, which is, however, considered in the following analysis.
    }
    \label{fig: kfold}
\end{figure*}

To evaluate the effects of the data partition on the performance of EEG-DL classifiers, 5 different cross-validation strategies were investigated.
They differentiate for the number of folds, the number of independent sets, and the way EEG windows are distributed across them.
CVs can be grouped in three main categories, schematized in \autoref{fig: kfold}, and concisely described below.

\WarningsOff
\begin{enumerate}[\textbullet]
    \item \textit{Sample-based K-Fold}: a standard K-Fold cross-validation, where windows of the same EEG recording can appear both in the training and in the validation set.
    While still used in recent applications \cite{pdcnnet}, this approach has the disadvantage of not providing accuracy estimates on unseen subjects.
    However, EEG signals are characterized by strong subject-specific characteristic, which can be learned during the training and leveraged by the models during inference, potentially generating optimistic results \cite{SubjAwSSL}.
    This study considers a 10-Fold sample-based cross-validation, which will be referred to as K-Fold. 
    \item \textit{Leave-N-Subjects-Out}: also called subject-based cross-validation.
    It is a variant of the sample-based K-Fold cross-validation where folds are created by randomly assigning EEG windows of the same subject either in the training or in the validation set; hence, it allows to validate the model only on samples from unseen subjects, providing more realistic accuracy estimates.
    However, this approach does not consider separate validation and test sets to respectively control the training and evaluate the model, which is crucial to train EEG-DL models that are notably known to be very susceptible to overfitting.
    Several papers have used the validation set to both monitor the training and evaluate the model; yet, this approach is not appropriate since it introduce data leakage, as described in \cite{nloso_emo}.
    This study considers a 10-Fold subject-based cross-validation, which will be referred from here as LNSO, and a Leave-One-Subject-Out cross-validation, where each fold is composed by the records of only one subject. From here, the Leave-One-Subject-Out cross-validation will be referred to as LOSO.
    \item \textit{Nested-Leave-N-Subjects-Out}: as proposed in \cite{DelPup2024b, nloso_emo}, introduces a nested level for the creation of multiple train-validation-test splits. 
    Each triplet is created by concatenating two different subject-based cross-validation (LOSO or LNSO). The first determines the subjects to use as test set, called outer folds in \autoref{fig: kfold}, while the second further partition the remaining subjects to create the training and validation sets (inner folds).
    In this way, a model can be trained, monitored, and evaluated on independent sets, removing any form of data leakage.
    However, the addition of a nested level increases the total number of training and computational time.
    Nested approaches result in a total of $N_{outer} \times N_{inner}$ training, which can reach the limit value of $N_{subj} \times (N_{subj}-1)$ in case of a fully Nested-Leave-One-Subject-Out cross-validation.
    This study considers a 10-outer/10-inner nested subject-based cross-validation (100 partitions), which will be referred to as N-LNSO, and a fully Nested-Leave-One-Subject-Out cross-validation, characterized by the concatenation of two LOSO, which will be referred to as N-LOSO.
    Test sets defined by outer folds perfectly overlap validation sets of standard subject-based CVs (LNSO or LOSO), which is crucial for the comparative analysis presented in subsections \ref{subsec: lnso vs nlnso} and \ref{subsec: loso vs nloso}.
\end{enumerate}
\WarningsOn

To summarize, Figure {\ref{fig: kfold}} illustrates three main categories of cross-validation, which differ in terms of the number of independent sets (nested vs. non-nested) and the distribution of the EEG windows (sample-based vs subject-based). Furthermore, the number of left-out subjects (N) is also an important cross-validation parameter. When N is set to 1, two additional settings emerge: Leave-One-Subject-Out (LOSO) and the Nested-Leave-One-Subject-Out (N-LOSO). In EEG deep learning studies, the Leave-One-Subject-Out approach is frequently used to assess the performance of deep neural models. Thus, it is crucial to consider it separately to evaluate its reliability. The same consideration applies for the Nested-Leave-One-Subject-Out method, which is considered separately from the Nested-Leave-N-Subjects-Out (N-LNSO).

Considering this, a total of CV five settings based on the 3 categories described above was used for the comparisons presented in Section {\ref{sec: results}}. Each of these settings has been assigned a unique acronym, which is summarized below:

\WarningsOff
\begin{enumerate}[\textbullet]
\item Sample-based K-Fold, called K-Fold.
\item Leave-N-Subjects-Out, called LNSO.
\item Leave-One-Subject-Out, called LOSO.
\item Nested-Leave-N-Subjects-Out, called N-LNSO.
\item Nested-Leave-One-Subject-Out, called N-LOSO.
\end{enumerate}
\WarningsOff

\subsection{Implementation details}
\label{subsec: implement}
The remaining steps of the experiment were implemented within a Python environment.
In particular, models were trained using \textit{SelfEEG}\footnote{https://github.com/MedMaxLab/selfEEG} \cite{DelPup2024c} (v0.2.0) and figures were generated with \textit{Seaborn} \cite{seaborn} using the Wong's colorblind palette.
Experiments were conducted mainly on the Padova Neuroscience Center computing node composed of two NVIDIA Tesla V100 GPU devices (CUDA 12.1), to speed up the entire process by running multiple training sessions in parallel.
The maximum GPU memory allocation was 8.01 GB.
Additionally, the large number of training instances of the N-LOSO (see Section A.1 of the Supplementary Materials) was completed with the addition of the Department of Neuroscience computing node, composed of three NVIDIA A30 GPU devices (CUDA 12.2).
Further details not reported in the following sub-paragraphs can be found in the supplementary materials and in the openly available source code.

\subsubsection{EEG Architectures}
\label{model-selc}
Four established EEG-DL models with increasing complexity (in terms of model size and framework, see {\cite{modelcompl}}) have been selected for analysis: EEGNet, ShallowConvNet, DeepConvNet, and Temporal-based 1D ResNet \cite{eegnet, shallow, resnet1}.

EEGNet is a compact deep neural model widely used for the tasks being investigated.
It consists of a series of convolutional layers that perform operations on either the temporal (sample) or spatial (channel) dimensions.
ShallowConvNet and DeepConvNet are two other architectures related to the EEGNet family {\cite{eegnetfam}}.
ShallowConvNet is a more compact version of EEGNet, while DeepConvNet includes additional convolutional layers in its encoder block.
Similarly, ResNet is a deeper architecture recently employed in several EEG-based classification tasks, thanks to recent advances in unsupervised pretraining strategies such as self-supervised learning \cite{rafieissl, resssl}.
Numerous variants of EEG-based ResNet models exist in the literature, primarily differing in the number of stacked residual blocks and the combination of temporal and spatial layers within those blocks \cite{resnet2}.
This study will use a variant of the ResNet34 model characterized by only temporal (horizontal) kernels, as proposed in \cite{resnet1}.
From this point forward, this model will be referred to as T-ResNet.

Model complexity was evaluated considering several factors, including the number of layers (architecture depth), the number of learnable parameters, and the required computational resources in terms of training time and GPU memory allocation.
Consequently, the number of parameters ranges from less than \num{100000} in ShallowConvNet configurations  to more than a \num{1000000} in T-ResNet ones.
A schematic representation of each model can be found in the supplementary materials, with further implementation details within the openly available source code.

\subsubsection{Training Hyperparameters}
\label{train-hyp}
To ensure consistent results in this study, a custom seed (83136297) was randomly selected and fixed.
This approach minimizes randomness in the code and enhances the reproducibility of results, as suggested in {\cite{DelPup2024a}}. 
Fixing the random seed affects the initialization of model weights and the creation of mini-batches. It also affects the subject assignment in the LNSO and N-LNSO cross-validation settings (LOSO and N-LOSO remain deterministic since each fold consists of a single subject).
Given the high variability of the results presented in Section {\ref{sec: results}}, an analysis of the effect of the random seed on cross-validation accuracy estimation is provided in Section A.2 of the Supplementary Materials.
This analysis aims to confirm the consistency of the findings of this study.

Models were initialized using Pytorch's \cite{pytorch} default settings, which varies according to the specific type of layer.
Models were trained with Adam optimizer ($\beta_{1} = 0.9$, $\beta_{2} = 0.999$, no weight decay) \cite{adam}, using batch size of 64 and cross entropy as loss function.
The maximum number of epochs was set to 100.
In addition, an early stopping was used to control the training and restore the model's best weights in case the validation loss did not improve for 15 epochs.
It is worth recalling that this approach produces biased results in standard cross-validation strategies, because they do not have separate validation and test sets.
Using the same set to both stop the training and evaluate the model put the parameter's optimization search in an optimistic environment, introducing a minor but still relevant form of data leakage.
This is a known but often ignored issue and it was purposely kept in the following analysis to evaluate its effect.
Further details on this matter are discussed in Section \ref{sec: discussion}.
Following the strategy described in {\cite{DelPup2024b}}, a custom learning rate was selected for each combination of task and model, as summarized in Table {\ref{tab: learningrate}}.
Learning rates were determined by evaluating the balanced accuracy of validation sets (considering both median and inter-quartile range) for a subset of models trained using the N-LNSO cross-validation scheme (see Figure {\ref{fig: kfold}}). This evaluation involved searching over a non-uniform discrete grid of 13 potential values:
$1.0\cdot10^{-3}$, $7.5\cdot10^{-4}$, $5.0\cdot10^{-4}$, $2.5\cdot10^{-4}$, $1.0\cdot10^{-5}$, $7.5\cdot10^{-5}$, $5.0\cdot10^{-5}$, $2.5\cdot10^{-5}$, $1.0\cdot10^{-6}$.
This procedure provides a good compromise between computational efficiency and the quality of the selected hyperparameters.
It also avoids bad practices, such as cross-validating the entire dataset, and minimizes potential selection bias caused by the high inter-subject variability, which can lead to suboptimal values when the process is conducted on a single hold-out data partition.
Furthermore, it does not favor any of the investigated cross-validation settings, as the N-LNSO validation accuracies are not considered in the comparative analysis.
Lastly, learning rates slightly decrease during training following an exponential scheduler with $\gamma = 0.995$.

\renewcommand{\arraystretch}{1.2} 
\begin{table}[!t]
\caption{Learning rate grid.}
\begin{tabular}{cccc}
\toprule
Model & BCI & Parkinson & Alzheimer \\
\midrule
ShallowConvNet & $7.5\cdot10^{-4}$ & $2.5\cdot10^{-4}$ & $5.0\cdot10^{-5}$ \\
EEGNet & $1.0\cdot10^{-3}$ & $1.0\cdot10^{-4}$ & $7.5\cdot10^{-4}$ \\
DeepConvNet & $7.5\cdot10^{-4}$ & $2.5\cdot10^{-4}$ & $7.5\cdot10^{-4}$ \\
T-ResNet & $5.0\cdot10^{-4}$ & $1.0\cdot10^{-5}$ & $5.0\cdot10^{-5}$ \\
\bottomrule
\end{tabular}
\label{tab: learningrate}
\end{table}
\renewcommand{\arraystretch}{1} 

\subsubsection{Performance Evaluation}  
\label{sec: performance}
The performance of each model was evaluated using the balanced accuracy, which can be defined as the macro average of recall scores per class \cite{metricsML}.
Hence, for a multi-class classification problem, with $K$ classes:
\begin{equation}
    \text{Accuracy}_{\textit{Balanced}} = \dfrac{1}{K}\sum_{i=1}^K\dfrac{\text{TP}_i}{\text{TP}_i + \text{FN}_i}
    \label{equ: BA_Nclasses}
\end{equation}
with $\text{TP}_i$ and $\text{FN}_i$ being respectively the true positives and the false negatives for class $i$.
Model performance against other metrics, such as the weighted F1-score, can be found in Section A.3 of the Supplementary Material, as well as in the openly available source code.

\section{Results}
\label{sec: results}

The analysis highlighted differences not only between sample-based and subject-based CV approaches, but also between nested and non-nested ones.
To organize the large volume of information gathered from the thousands of trained models, several figures are presented and described in three separate sub-paragraphs.
Each sub-paragraph compares two cross-validation approaches, namely:
\WarningsOff
\begin{enumerate}[\textbullet]
    \item LNSO vs. K-Fold (subject-based vs Sample-based CV)
    \item LNSO vs. N-LNSO
    \item LOSO vs. N-LOSO
\end{enumerate}
\WarningsOn
and aims at answering the following questions:
\WarningsOff
\begin{enumerate}[\textbullet]
    \item Are there differences between sample-based and subject-based CVs?
    \item Are there differences between nested and non-nested subject-based CVs?
    \item Are variations in the accuracies between nested and non-nested subject-based CV strategies influenced by model complexity?
\end{enumerate}
\WarningsOn

\begin{figure*}[!t]
    \centering
    \includegraphics[width=0.95\textwidth]{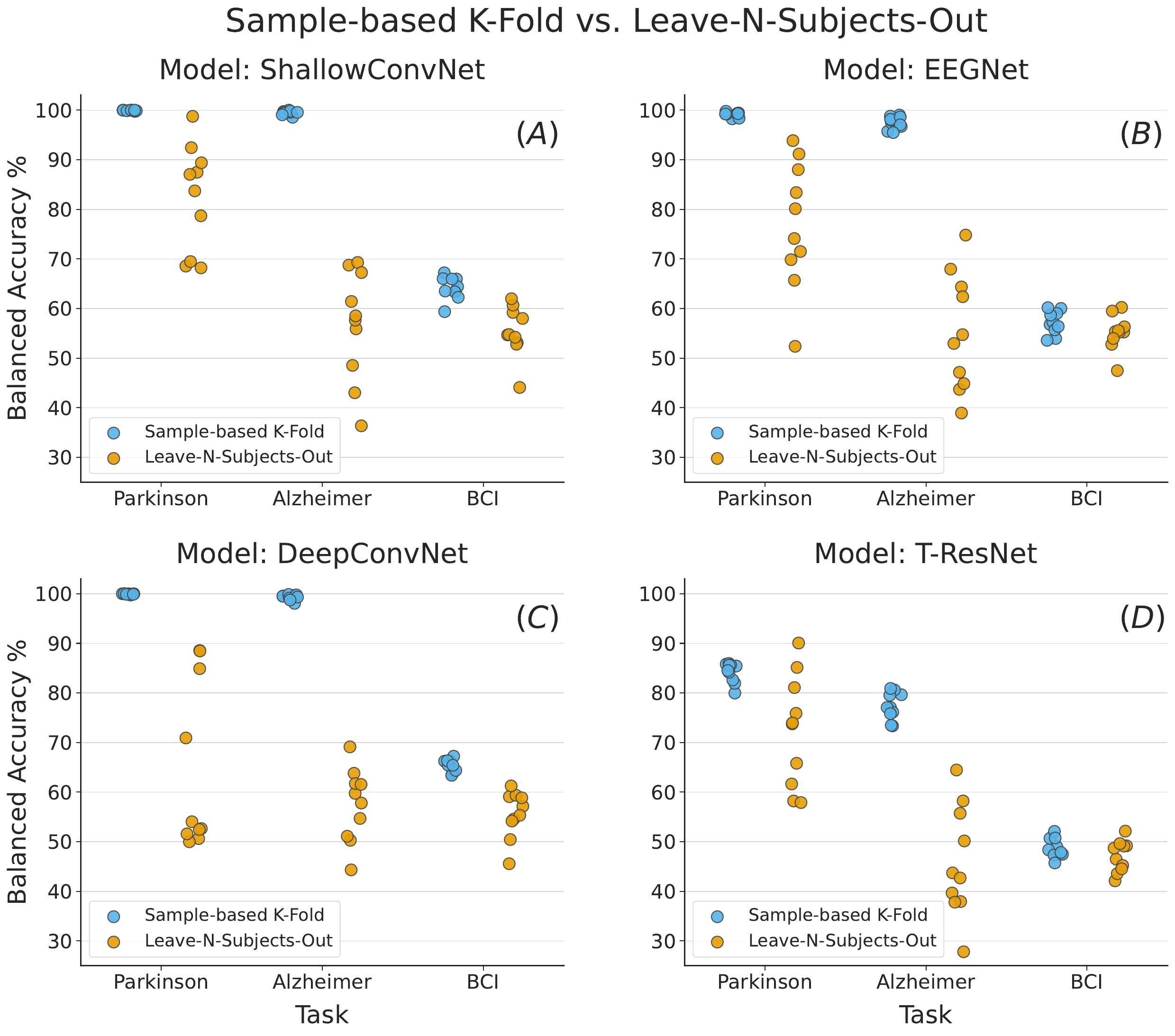}
    \caption{
    Balanced accuracy comparison between Sample-based (K-Fold, in blue) and subject-based (LNSO, in orange) 10-Fold cross-validations.
    The sub-figures display results for different deep learning architectures: ShallowConvNet (Panel A), EEGNet (Panel B), DeepConvNet (Panel C), and T-ResNet (Panel D).
    Results are shown across all tasks.
    Independently from the model, there is a performance drop and an increased variance when switching to subject-based cross-validation methods, particularly in pathology classification tasks.
    Additionally, ShallowConvNet, the smallest model, achieved the highest median accuracies, highlighting a potential drawback in the usage of more complex models.}
    \label{fig: kfoldvslnso}
\end{figure*}
\begin{figure*}[!t]
    \centering
    \includegraphics[width=0.95\textwidth]{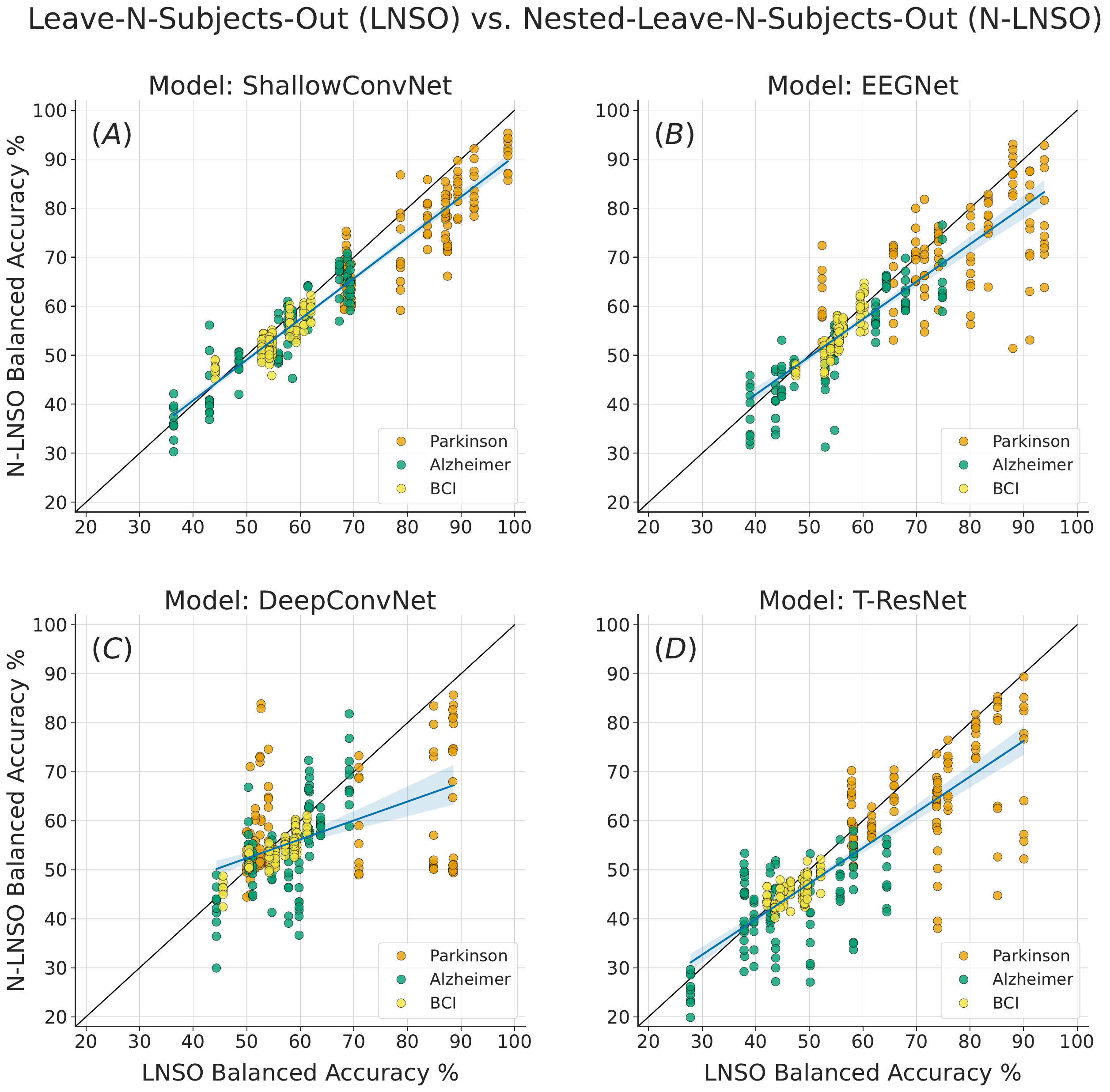}
    \caption{
    Comparison of balanced accuracy between Nested-Leave-N-Subjects-Out (N-LNSO) and Leave-N-Subjects-Out (LNSO). The sub-figures display results for different deep learning architectures: ShallowConvNet (Panel A), EEGNet (Panel B), DeepConvNet (Panel C), and T-ResNet (Panel D). Results are shown across all tasks, with light blue representing Parkinson's, green for Alzheimer's, and yellow for BCI.
    Each column of points shows N-LNSO results for architectures trained on different train/validation partitions (inner folds) but evaluated on the same test set (outer fold), as a function of LNSO accuracies.
    Notably, performance differences between the two CVs increase with higher LNSO fold accuracies.
    The regression lines' slopes (in red), always less than 1, further highlight this trend.
    Additionally, more complex models exhibit greater result variance.
    }
    \label{fig: lnsovsnlnso}
\end{figure*}

The third question is particularly relevant, because it addresses the current trend of developing larger and deeper EEG-DL models without increasing the amount of data on which they are trained (e.g., {\cite{dicenet}}).
In supervised deep learning, overparameterization has been found to improve the optimization landscape of a problem, often with minimal impact on generalization {\cite{overparametrization}}. However, the effects on generalizability may be more pronounced when dealing with highly complex and noisy data, like EEG signals. Additionally, the limitations of traditional cross-validation methods may obscure these effects, leading to an overestimation of model generalizability and potentially favoring overparameterized models over shallow models.
Therefore, this study aims to investigate whether EEG-DL models take advantage of limitations of non-nested cross-validation methods---such as data leakage and improper usage of the validation set---to the same extent.

\subsection{Sample-based vs subject-based cross-validation}
\label{subsec: kfold vs lnso}

Figure \ref{fig: kfoldvslnso} presents a comparative analysis of LNSO and sample-based K-Fold results.
Regardless of the model, K-Fold results consistently outperform those of the LNSO, particularly in pathologic tasks, where balanced accuracies of ShallowConvNet (Panel A), EEGNet (Panel B), and DeepConvNet (Panel C) almost reach 100\%.
Furthermore, the variance of results shows significant divergence between the two cross-validation methods.
In particular, differences between the minimum and maximum balanced accuracies often exceed 30\% across all the four models, particularly in the Parkinson’s and Alzheimer’s disease classification tasks.
Similar trends are observed in the BCI results both in terms of performance bias and variance, albeit to a much lower degree.
Smaller architectures yield higher performances in subject-based cross-validation. 
Specifically, ShallowConvNet achieves median balanced accuracies of 85.4\% for Parkinson's task, 58.1\% for Alzheimer's, and 54.7\% for BCI.
In contrast, worst performances are registered by DeepConvNet for Parkinson (53.3\%), and by T-ResNet (Panel D) in both Alzheimer's (43.2\%) and BCI (47.6\%) tasks.
These results suggest potential limitations associated with deploying larger, non-pretrained deep learning models.

\subsection{Leave-N-Subjects-Out vs Nested-Leave-N-Subjects-Out}
\label{subsec: lnso vs nlnso}

Figure \ref{fig: lnsovsnlnso} illustrates the results of the N-LNSO in relation to those from the LNSO.
As noted in subsection \ref{subsec: CV}, this comparison is enabled by the complete overlap between the test sets of the N-LNSO and the validation sets of the LNSO.
In particular, each set of N-LNSO accuracies (column of points) having the same x-axis value---which is equal to the LNSO accuracy calculated from the partition using as validation set the identical group of subjects used as test set in the N-LNSO---highlights how model performance is influenced by the definition of the validation set.
Focusing on the model architecture, a relationship emerges between the variability of N-LNSO results and model complexity.
ShallowConvNet (Panel A) appears less sensitive to changes in the training/validation partition (inner folds), resulting in differences between the minimum and maximum balanced accuracies that typically remain below 20\%.
The same can be observed in EEGNet (Panel B) for the Alzheimer's task.
In contrast, both DeepConvNet (Panel C) and T-ResNet (Panel D) exhibit greater performance variability, with T-ResNet displaying accuracy differences exceeding 40\% in the Parkinson's task. 
Furthermore, median N-LNSO test accuracies derived from groups of partitions sharing the same test set (i.e., all inner folds corresponding to a specific outer fold, as shown in Figure \ref{fig: kfold}-C) often fall below the bisector.
This observation might support the hypothesis that traditional subject-based cross-validations tend to produce inflated results due to the lack of independent validation and test sets, as noted in \cite{nloso_emo}.
Specifically, since N-LNSO combines two 10-Fold subject-based CVs, we can identify 30 groups (10 for each task) of 10 accuracies derived from the same test set, indicated by the columns of points in Figure \ref{fig: lnsovsnlnso}).
Among those 30 groups, only 5, 4, 9, and 8, exhibit median balanced accuracies exceeding those of the LNSO for EEGNet, ShallowConvNet, DeepConvNet, and T-ResNet, respectively. 
Additionally, differences in balanced accuracy tend to increase with higher LNSO accuracies, a trend supported by the slope of the regression lines showed in the same figure.
However, ShallowConvNet, being the simplest model, appears less affected by this trend, as confirmed by its regression slope of 0.83 ($R^2_{\text{adj}}=0.91$), which is the closest to 1.

\subsection{Leave-One-Subject-Out vs Nested-Leave-One-Subject-Out}
\label{subsec: loso vs nloso}

\begin{figure*}[p]
    \centering
    \includegraphics[width=0.99\linewidth]{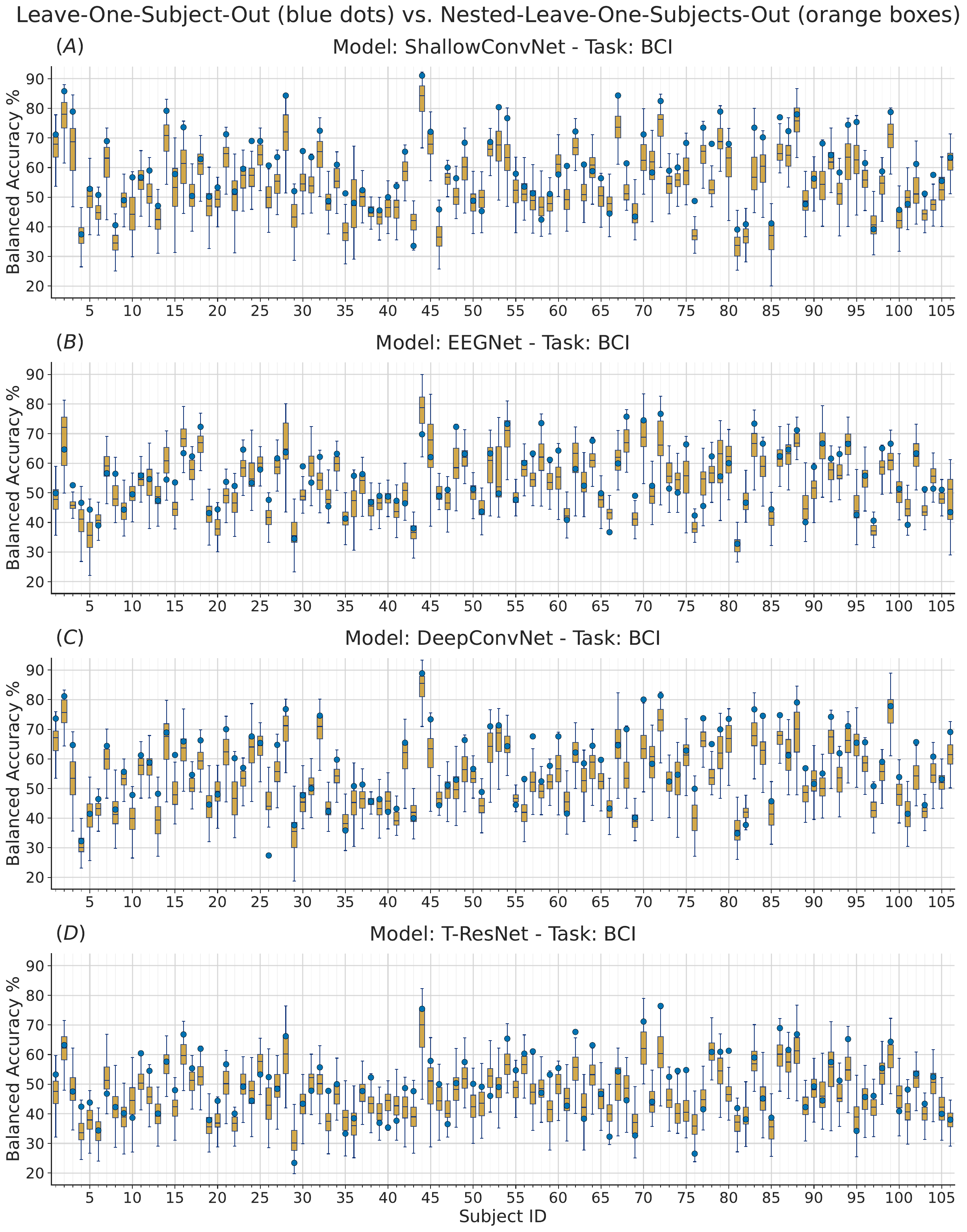}
    \caption{Balanced accuracy comparison between Nested-Leave-One-Subject-Out (N-LOSO) and Leave-One-Subject-Out (LOSO) for the BCI task. Each blue dot reports the subject-wise LOSO balanced accuracy resulted from a model trained on all the subjects except one left-out used as validation set. Each boxplot gathers all the balanced accuracies of models evaluated on the same left-out subject, but trained on different train/validation partitions following a LOSO scheme. In other words, each model was trained on all except two subjects, monitored on another changed every training, and evaluated on the same left-out subject.}
    \label{fig:nlosobci}
\end{figure*}

Figures \ref{fig:nlosoalz}, \ref{fig:nlosopd}, and \ref{fig:nlosobci} provide a comparative analysis of the results obtained from LOSO and N-LOSO results.
As highlighted in the previous comparison, the absence of independent validation and test sets might explain the noticeable increase in the LOSO performance metrics.
Specifically, Table \ref{tab:losowin} summarizes the proportion of subjects for which the median N-LOSO test accuracy outperform that of the LOSO.
This proportion remains below 10\% for all the models in pathologic tasks and below 36\% for the BCI task, although in the latter scenario, this increase may be attributed to the inherent characteristic of the task, as noted in subsection \ref{subsec: kfold vs lnso} and elaborated in section \ref{sec: discussion}.
Moreover, table \ref{tab:losowin} compares median performance metrics alongside their interquartile ranges.
Notably, LOSO results suggest a superiority of DeepConvNet in pathology classification tasks (ShallowConvNet exhibit comparable performance in BCI), with a median balanced accuracy of respectively 100.00\% in Parkinson's task and 90.62\% in Alzheimer's.
However, when the evaluation is shifted to a nested scenario, it is ShallowConvNet, the smallest architecture, that achieves the highest performance, with a lower but still remarkable percentage change of median values.

Figures \ref{fig:nlosoalz}, \ref{fig:nlosopd}, and \ref{fig:nlosobci} also highlight the dramatic subject-wise variability observed in N-LOSO results, despite training sets across inner folds differing by only a single left-out subject used for validation purposes. 
In the BCI task, the range between minimum and maximum balanced accuracies fluctuates between 20\% and 30\%; however, for the pathological tasks, these values exhibit a remarkable increase, sometimes encompassing the entirety of the accuracy spectrum.

\begin{figure*}[!t]
    \centering
    \includegraphics[width=0.91\textwidth]{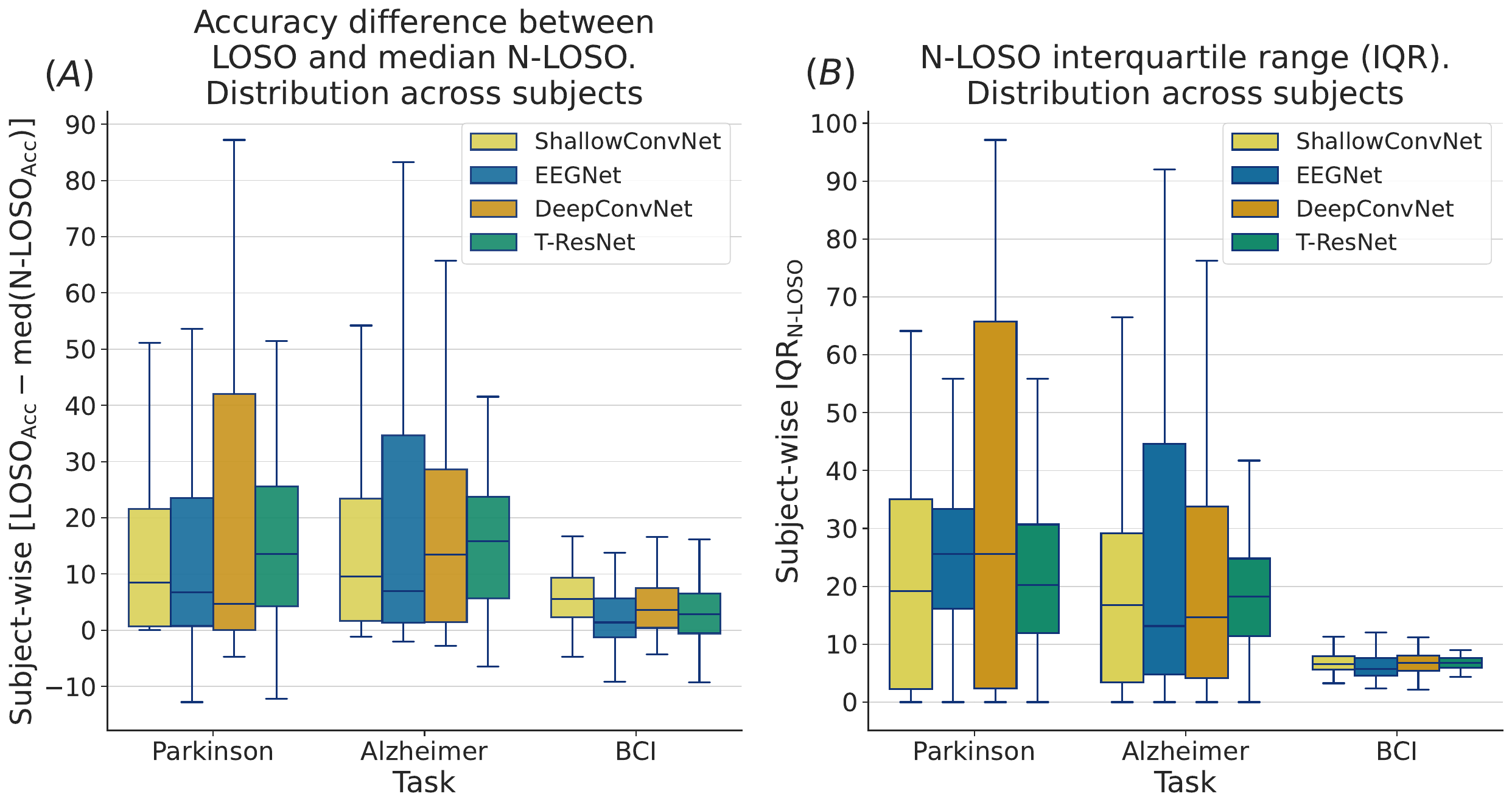}
    \caption{
    Subject-wise analysis of Nested-Leave-One-Subject-Out (N-LOSO) results. On the left, the distribution of differences across all subjects between the Leave-One-Subject-Out (LOSO) balanced accuracies and the median N-LOSO is evaluated for each model architecture and classification task.
    On the right, the interquartile range distribution of N-LOSO results for each subject is evaluated for the same architectures and classification tasks. Both figures highlight minimal differences in the BCI task, but reveal higher performance drops in pathologic tasks for more complex models such as T-ResNet.
    }
    \label{fig: losovsnlosoiqr}
\end{figure*}

Figure \ref{fig: losovsnlosoiqr} further investigates possible relations between model complexity and performance declines in N-LOSO, presenting the subject-wise distributions of performance bias between LOSO and N-LOSO (Panel A), calculated as the median accuracy across all inner folds, alongside the interquartile ranges of N-LOSO (Panel B).  
Even in this case, no notable differences can be found when looking at BCI results.
However, in both clinical applications, T-ResNet (the largest model) distributions are centered at higher values compared to ShallowConvNet (the simplest model).
Furthermore, the distributions of DeepConvNet showed greater variability, particularly for the Parkinson's task, with upper whiskers reaching higher values. 

\renewcommand{\arraystretch}{1.2} 
\begin{table*}[!t]
\caption{Comparative analysis between LOSO and N-LOSO balanced accuracies. The percentage changes indicate the amount of increase or decrease the N-LOSO balanced accuracy median and IQR values have relatively to the LOSO metrics, in percentage.}
\begin{tabular}{cccccccc}
\toprule
\multirow{2}{*}{Task} & \multirow{2}{*}{Model} & \multicolumn{2}{c}{\makecell{Balanced Accuracy\\Median $[25^{th}$$-75^{th}]$}} & \multicolumn{2}{c}{\makecell{Balanced Accuracy\\Percentage Change\\N-LOSO vs LOSO}} & \multirow{2}{*}{\makecell{Number\\of subjects\\N-LOSO > LOSO}} \\
 &  & LOSO & N-LOSO & Median & IQR &  \\
\midrule
& ShallowNet & \makecell{99.38 \\$[80.95-100.00]$} & \makecell{88.37\\$[50.00-100.00]$} & \makecell{-11.08} & +162.50 & 0/81 \\
\multirow{2}{*}{Parkinson} & EEGNet & \makecell{97.83 \\$[82.61-100.00]$} & \makecell{76.68\\$[42.97-97.81]$} & \makecell{-21.62} & +215.36 & 2/81 \\
& DeepConvNet & \makecell{100.00\\$[89.63-100.00]$} & \makecell{81.67\\$[10.80-100.00]$} & \makecell{-18.33} & +760.51 & 1/81 \\
& T-ResNet & \makecell{91.36\\$[75.00-97.67]$} & \makecell{70.19\\$[35.71-91.50]$} & \makecell{-23.17} & +146.04 & 2/81 \\
\midrule
& ShallowNet & \makecell{88.19\\$[48.64-97.34]$} & \makecell{63.05\\$[10.58-91.54]$} & \makecell{-28.51} & +66.23 & 4/88 \\
\multirow{2}{*}{Alzheimer} & EEGNet & \makecell{94.42 \\$[48.36-98.53]$} & \makecell{59.95\\$[3.65-94.8]$} & \makecell{-36.51} & +81.70 & 9/88 \\
& DeepConvNet & \makecell{90.62\\$[36.19-99.00]$} & \makecell{52.44\\$[5.44-93.12]$} & \makecell{-42.13} & +39.59 & 6/88 \\
& T-ResNet & \makecell{73.94\\$[32.43-93.67]$} & \makecell{45.15\\$[14.40-77.85]$} & \makecell{-38.94} & +3.62 & 7/88 \\
\midrule
& ShallowNet & \makecell{58.97\\$[50.83-68.96]$} & \makecell{53.56\\$[46.58-62.25]$} & \makecell{-9.18} & -13.54 & 14/106 \\
\multirow{2}{*}{BCI} & EEGNet & \makecell{54.09 \\$[47.48-63.18]$} & \makecell{52.28\\$[45.24-60.16]$} & \makecell{-3.34} & +5.00 & 39/106 \\
& DeepConvNet & \makecell{58.90\\$[48.04-67.28]$} & \makecell{52.38\\$[44.57-62.04]$} & \makecell{-11.06} & -9.19 & 19/106 \\
& T-ResNet & \makecell{48.85\\$[42.29-56.46]$} & \makecell{46.00\\$[39.97-53.06]$} & \makecell{-5.84} & -7.59 & 31/106 \\
\bottomrule
\end{tabular}
\label{tab:losowin}
\end{table*}
\renewcommand{\arraystretch}{1} 

\section{Discussion}
\label{sec: discussion}

Developing deep learning pipelines for EEG applications is a complex task.
Minimal variations in the experimental setup can lead to significant variations in the results, with data partitioning being particularly influential.
In this field, the presence of datasets containing only a few dozen subjects does not allow the use of hold-out strategies, as model performance is highly dependent to the composition of the test set.
For instance, the test set accuracies derived from individual splits of the N-LNSO, illustrated in figure \ref{fig: lnsovsnlnso}, demonstrate how specific partitions can favor one architecture over another.
Therefore, it is crucial to assess model performance using cross-validation strategies.
However, even cross-validation procedures must address the high inter-subject variability inherent in EEG signals, as ignoring this can lead to unrealistic performance evaluations.

The comparative analysis in subsection \ref{subsec: kfold vs lnso} highlights that sample-based methods, which allow windows of the same EEG signal to appear both in the training and validation sets, do not satisfy this requirement.
Sample-based methods fail to evaluate the model's ability to generalize to entirely unseen subjects and overlook potential relationships between consecutive windows, even when they are temporally separated, potentially inflating performance results.
Consequently, it is not surprising to observe significant drops in performance during the transition to subject-based methods, particularly in applications where each subject has a unique label reflecting their health status.
The results in Subsection {\ref{subsec: kfold vs lnso}} offer a new perspective on the exceptionally high performance reported in recent scientific papers applying deep learning to EEG data. They suggest that such performances might stem more from the propensity of deep neural networks to overfit due to the inherent characteristics of EEG signals. These characteristics include subject dependence induced by biometric properties of the signal {\cite{tatar}}, inter-correlation of consecutive samples {\cite{sample_leakage_2}}, and other factors such as non-stationarity {\cite{kamrudloso}}.
The nearly perfect accuracies reported in many studies of the domain, such as \cite{pdcnnet, PD_oh, PD_Shaban, nosubj1, nosubj2}, may not necessarily demonstrate that such deep learning models effectively capture the representative features of the condition under investigation.
Instead, these accuracies may better reflect the models' ability to recognize individual subjects by their EEG signals \cite{tatar, sample_leakage_2}.

\begin{figure*}[p]
    \centering
    \includegraphics[width=0.99\linewidth]{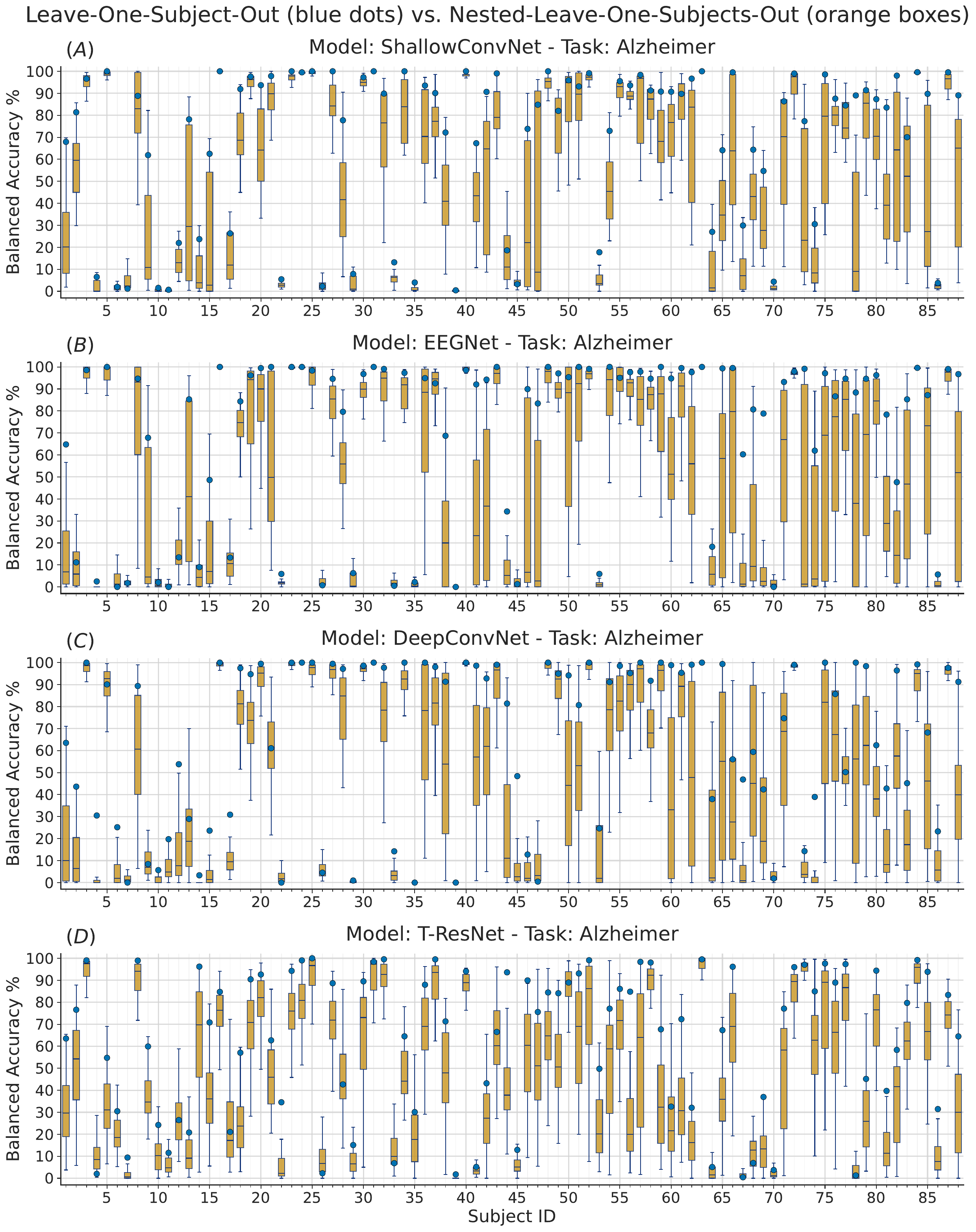}
    \caption{Balanced accuracy comparison between Nested-Leave-One-Subject-Out (N-LOSO) and Leave-One-Subject-Out (LOSO) for the Alzheimer's task. Each blue dot reports the subject-wise LOSO balanced accuracy resulted from a model trained on all the subjects except one left-out used as validation set. Each boxplot gathers all the balanced accuracies of models evaluated on the same left-out subject, but trained on different train/validation partitions following a LOSO scheme. In other words, each model was trained on all except two subjects, monitored on another changed every training, and evaluated on the same left-out subject.}
    \label{fig:nlosoalz}
\end{figure*}

\begin{figure*}[p]
    \centering
    \includegraphics[width=0.99\linewidth]{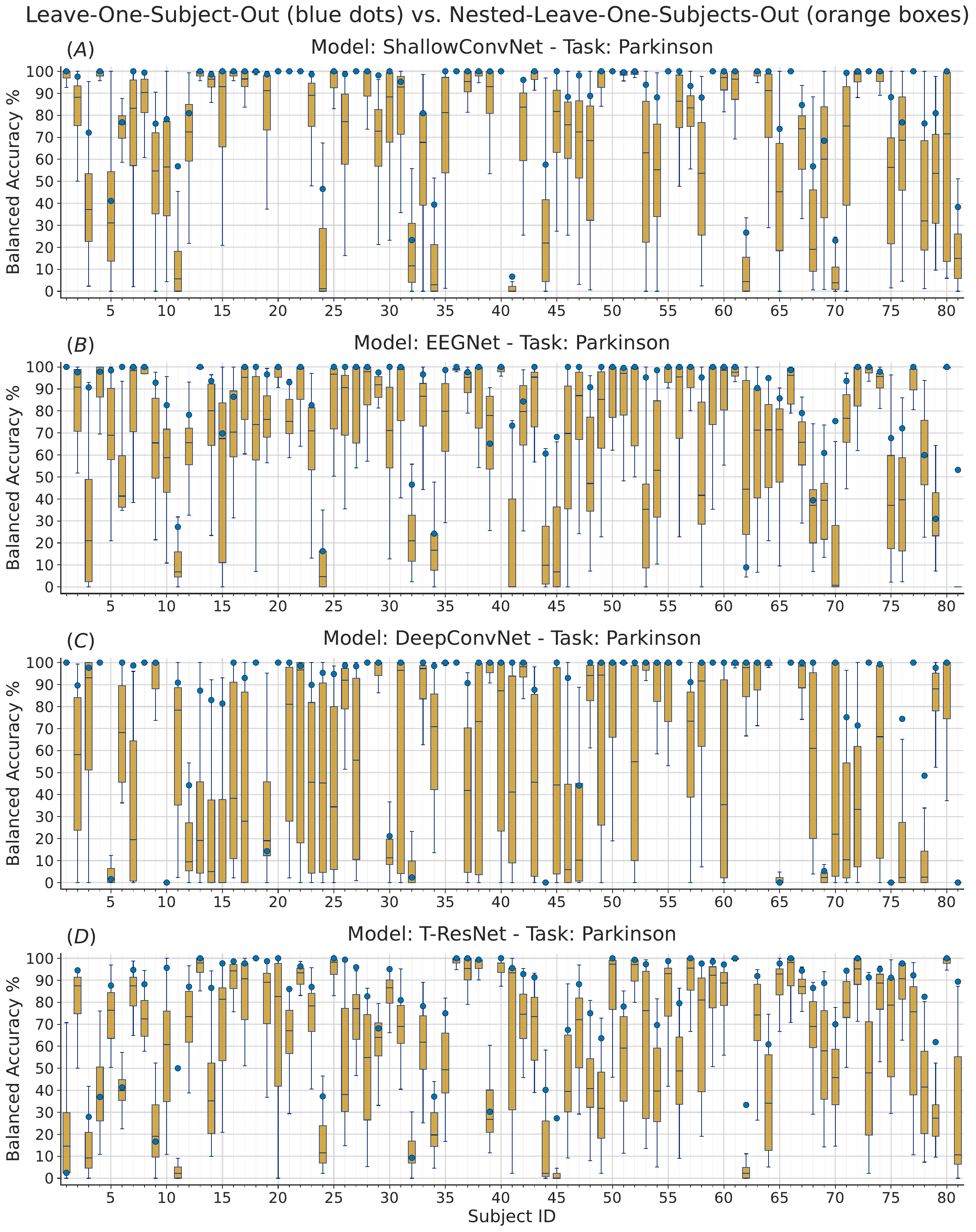}
    \caption{Balanced accuracy comparison between Nested-Leave-One-Subject-Out (N-LOSO) and Leave-One-Subject-Out (LOSO) for the Parkinson's task. Each blue dot reports the subject-wise LOSO balanced accuracy resulted from a model trained on all the subjects except one left-out used as validation set. Each boxplot gathers all the balanced accuracies of models evaluated on the same left-out subject, but trained on different train/validation partitions following a LOSO scheme. In other words, each model was trained on all except two subjects, monitored on another changed every training, and evaluated on the same left-out subject.}
    \label{fig:nlosopd}
\end{figure*}

Creating subject-specific representations is not equivalent to creating label-specific representations.
If a model is overfitted to specific subjects, it will likely assign labels based on similarity to those seen during the training phase.
Thus, the performance of the model is closely related to the degree of similarity between the training, validation and test sets, or more specifically, to how similar subjects with the same label (or health status) are to each other.
This interpretation not only explains the observed variability in the folds of the LNSO but also clarifies why performance in the N-LOSO setting can be closed to  zero in some cases (Figures {\ref{fig:nlosoalz}} and {\ref{fig:nlosopd}}).
Performance metrics close to zero indicate that the model predictions are consistently incorrect. This situation is different from that of a random classifier, as it suggests that a decision rule has been learned, but it is ineffective for solving the task.
This result raises fundamental concerns about the features learned by EEG deep learning models, underscoring the need to improve their interpretability in this domain.

The dominance of subject-specific characteristics in the learned embeddings may also explain why performance on the BCI task is generally lower and with less variance.
In the BCI task, each subject participated in trials of every possible motor/imagery movement, preventing the model from exploiting subject-specific characteristics as shortcuts to minimize the training loss.
This may explain why reported performances, even in sample-based cross-validation methods, are relatively low, despite being above the baseline of random guessing classifiers (25\% accuracy for a four-class classification problem).
In addition, the selected dataset does not provide any information about the participants' level of preparation for executing the trials, especially those involving motor imagery.
The absence of such information complicates the identification of bad-trials to discard, contributing to another source of under-performance to consider during the analysis of the results.
While this study focuses on cross-subject prediction, the BCI field---and related EEG domains---includes real-world applications where it is acceptable to perform within-subject EEG data analysis by training subject-specific models.
However, it is necessary to acknowledge the limitations posed by the high inter-subject variability and the properties of the learned embeddings, which may focus on dominant subject-specific features that are not necessarily representative of the task of interest in general. Understanding these limitations can lead to different interpretations of study results and promote research in critical areas such as model interpretability and generalizability.
Enhancing model interpretability is crucial for understanding the features learned by EEG deep learning models, while improving model generalizability is essential for creating robust models applicable to real-world scenarios.
This approach can help make models less vulnerable to performance drops that may arise from changes in the subject’s health or psychological state, which may alter their normal brain activity.

An important issue identified in this study is the lack of independent validation and test sets in traditional subject-based approaches (LNSO and LOSO). 
This absence can lead to data leakage or overfitting.
To address this problem, the literature proposed two approaches.
The first one involves using the validation set to both monitor training and evaluate the model, rather than creating a third independent set from a subset of training samples (e.g., {\cite{noval1, noval2, noval3}}).
This method is incorrect because training stops when the model parameters reach the highest possible validation accuracy.
Data leakage occurs because validation data (which are also used to evaluate the model) are considered during training. Consequently, the model tends to overfit on the validation set, since it is used by the early stopping callback.

The second approach consists in creating a third independent set from a subset of training data to monitor model performance during training (e.g., {\cite{onlyoneval}}).
This method enhances methodological rigor by ensuring that the left-out subjects remain unseen during training and are used solely to evaluate the trained model.
However, it does not address how the selection of data for this additional set can influence model performance, potentially introducing a subtler form of result manipulation.
Figures {\ref{fig:nlosoalz}}, {\ref{fig:nlosopd}}, and {\ref{fig:nlosobci}} illustrate that specific training/validation/test partitions can yield higher performance than those from traditional subject-based cross-validations.
This finding reveals that it is possible to identify an optimal training/validation partition (inner fold) for each left-out subject (outer fold), resulting in a performance that exceeds that of a LOSO in terms of median value and interquartile range.
Following this strategy, DeepConvNet would achieve median balanced accuracies of 100\% (IQR: 0.6), 99\% (IQR: 30), and 64\% (IQR: 17) for the Parkinson's, Alzheimer's, and BCI tasks, respectively.

These results highlight the importance of using nested approaches in deep learning-based EEG data analysis to prevent overestimating model accuracy, which could otherwise occur under the guise of addressing the previously described issues.
They also emphasize the need for developing novel software tools to standardize EEG data and create large-scale multicenter datasets. 
Such efforts could help mitigate the discrepancies between model complexity and the amount of training data.
Given that the number of subjects is a primary source of variability, the release of large-scale datasets could significantly improve model generalizability and reduce the overfitting problems described above.
Increasing the number of large-scale datasets would facilitate the effective exploration of promising DL architectures, such as transformers.

Nested methods provide more realistic accuracy estimates that address the aforementioned data leakage issues, though they require greater computational resources.
This is important in deep learning-based EEG data analysis, where overparameterization can increase overfitting tendencies and early stopping serves as a key regularization technique (see Section A.6 of the Supplementary Materials).

\subsection{Guidelines for proper EEG model evaluation}

This study highlights the relationship between data partitioning in deep learning and EEG, emphasizing the importance of selecting the appropriate partitioning strategy based on the specific research task.
While data partitioning is a general concept in machine learning, its impact is especially pronounced in EEG due to the specific characteristics of the signals, such as their biometric features, the inter-correlation of consecutive samples, and factors like non-stationarity.
These properties can cause the learning phase to focus on extracting features that are useful for minimizing the loss but not relevant to the task at hand.

For research tasks targeting within-subject analysis via sample-based approaches, such as BCI experiments, models might be tailored to specific subject to enhance human-machine interaction for the potential end user.
Still, it is crucial to interpret the results carefully to avoid overstating claims.
In contrast, research tasks like disease prediction require cross-subject analyses, as sample-based methods heavily overestimate model generalizability, as discussed in subsection {\ref{subsec: kfold vs lnso}}. 
Choosing an inappropriate model evaluation approach may result in inflated and unrealistic performance estimates, which can limit the usefulness of the results in real-world scenarios.

Given the computational burden and the number of training instances, we suggest the following potential guidelines for selecting the most appropriate combination of outer and inner folds:
\begin{enumerate}
    \item \textit{Datasets with 20 or fewer subjects}: full N-LOSO ($N_{subj}\times(N_{subj}-1)$ training instances, maximum 380).
    \item \textit{Datasets with up to 50 subjects}: LOSO-outer/10-inner nested CV ($N_{subj}\times10$ training instances, maximum 500). In this case, each outer fold used as a test set will consist of a single left-out subject, while the inner folds will create validation sets comprising 10\% of the remaining subjects.
    \item \textit{Datasets with more than 50 subjects}: 10-outer/10-inner N-LNSO (always 100 training instances).
\end{enumerate}

These recommendations balance computational costs with the effects of the number of folds in terms of the bias-variance trade-off.
In point 1, the small number of subjects allows for computational feasibility with N-LOSO, and the increased variance in the results is justified by the improved performance estimation given the limited amount of data.
In point 2, while the number of subjects is still insufficient to produce representative test sets for conducting an N-LNSO, it is too large for a N-LOSO. 
A practical solution is to reduce the number of training instances by replacing the inner LOSO with a 10-fold LNSO.
The recommendation against reversing the order (using 10 outer folds and LOSO inner) is based on the importance of early stopping as a key regularization technique in EEG deep learning applications.
It is preferable to select the best weights based on a group of subjects to foster generalization to new subjects, rather than relying on a single subject that may be overfitted during training.
In point 3, the number of subjects becomes too large to justify the computational burden of running an N-LOSO cross-validation.
In contrast, the N-LNSO provides an optimal compromise in terms of the bias-variance trade-off.
The choice of ten for both inner and outer folds aligns with common practices and helps assess the model's stability against variation in the validation set.

\subsection{Limitations and future work}

Although the figures and tables strongly support the argument previously presented, it is essential to acknowledge two main limitations of the study.

First, no statistical tests have been used to analyze the results.
This decision is based on the violation of the assumptions required by both parametric and non-parametric statistic tests, specifically concerning the normality of the distribution of results and the independence of observations in this study.
The violation of sample independence is particularly relevant, as different partitions of the same cross-validation share overlapping training sets, especially as the number of folds increases.
This overlap can lead to gross underestimation of the variance in the cross-validation estimator and increase the risk of Type I errors.
Although previous studies \cite{Dietterich, BengioInf} have proposed modified versions of the t-test for single or repeated cross-validations that yield more conservative inferences, these tests still assume that accuracies from the folds follow a Gaussian distribution.
Furthermore, t-tests for cross-validation are intended to compare different classifiers within the same cross-validation, rather than comparing two cross-validations on the same classifier.
The use of non-parametric tests is also inappropriate, as both the independence of observations within the same CV and the lack of fully paired observations are violated. 
Therefore, it is more rigorous to provide a thorough quantitative and objective assessment of the results rather than overlook these violations just to produce optimistic and worthless p-values.

Secondly, while the analysis reveals strong differences among various cross-validation methods, it does not clarify whether these differences result from an overestimation of the generalization error by one method or an underestimation by another.
This question can be further investigated by considering the following point.
Throughout the discussion, it has been repeatedly emphasized how traditional cross-validation methods, when applied to EEG data, can introduce data leakage due to correlations between windows from the same signal and the manner in which training is monitored.
As a result, these procedures are naturally inclined to provide optimistic results.
To investigate and validate this hypothesis, a possible strategy may consist in adding a further nesting level to each of the investigated cross-validation methods in order to provide repeated measurements of the bias in the estimation of the generalization error.
However, this proposal is computational demanding and still affected by the high inter-subject variability of EEG data that can heavily affect each measurements.
Nevertheless, a preliminary analysis for the K-Fold vs LNSO and LNSO vs N-LNSO scenarios, with further details on its implementation, is provided in section A.4 of the Supplementary Materials.
This analysis confirms that both traditional cross-validation approaches tend to underestimate the generalization error and provide optimistic accuracies.

This study offers a rigorous analysis to the issues of overfitting and performance variability caused by data partitioning in EEG-based deep learning applications.
However, future research should validate its findings in different experimental contexts.
EEG data analysis using deep learning has a wide range of applications, including event-related potentials (ERPs), emotion and attention recognition, and sleep staging, to name a few.
It also involves various training paradigms (e.g., transfer, learning, few-shot learning, self-supervised learning), and levels of analysis (e.g., within, session, multi-center) that were not addressed in this study.
Moreover, the variability in performance of EEG deep learning systems is influenced by several factors beyond data partitioning, such as data harmonization, preprocessing strategies {\cite{DelPup2024b}}, and random seed control.
These factors deserve dedicated investigation in future works.


\clearpage

\section{Conclusion}
\label{sec: conclusion}

This study investigated the impact of data partition on the performance assessment of EEG-DL models.
Five distinct cross-validation strategies that operate either at the sample or at subject level are compared across three representative clinical and non-clinical classification tasks, using four established deep learning architectures with increased complexity.
The analysis of more than \num{100000} different trained models revealed strong differences between sample-based and subject-based approaches (e.g., Leave-N-Subjects-Out), highlighting how subject-specific characteristics can be learned and leveraged during inference to inflate performance estimates.
Such findings confirm the necessity of using subject-based strategies, particularly in clinical applications, where subject IDs and health status are uniquely identified.

Additionally, the analysis stressed the importance of maintaining independent validation and test sets to respectively monitor the training and evaluating the model.
Consequently, Nested-Leave-N-Subjects-Out (N-LNSO) was found to be sole method capable of preventing data leakage and providing more accurate estimation of model performance while accounting for the high inter-subject variability inherent to EEG signals.
In particular, a comparative evaluation of the Nested-Leave-One-Subject-Out (N-LOSO) and the Leave-One-Subject-Out (LOSO) cross-validation techniques demonstrated that traditional approaches tend to yield optimistic results.
Furthermore, it revealed that larger models exhibit higher performance drops as well as higher variance of results.
In summary, this study is the first to provide a comprehensive comparative analysis of different cross-validation methods for evaluating EEG deep learning models. It is the first to provide an analysis that goes beyond the sample-based vs. subject-based strategies (subsection {\ref{subsec: kfold vs lnso}}), clearly assessing the limitations of commonly used approaches, such as LNSO (subsection {\ref{subsec: lnso vs nlnso}}) or LOSO (subsection {\ref{subsec: loso vs nloso}}), and highlighting the importance of using nested strategies to provide more reliable performance estimates.
\printcredits

\section*{Declaration of competing interest}
The authors declare that they have no known competing financial interests or personal relationships that could have appeared to influence the work reported in this paper.

\section*{Code and data availability}
The code used to produce both results and figures is openly available at \href{https://github.com/MedMaxLab/eegpartition}{https://github.com/MedMaxLab/eeg-partition}.
All data that support the findings of this study are openly available within the OpenNeuro platform.

\section*{acknowledgment}
FDP would like to thank the Padova Neuroscience Center and its members for the support provided during the realization of this study.

\bibliographystyle{elsarticle-num}
\bibliography{bibliography.bib}

\clearpage
\setcounter{table}{0}
\setcounter{figure}{0}
\appendix

\section{Supplementary Materials}

The following subsections provides supplementary materials for the research study ``Rethinking how to evaluate model performance in deep learning-based cross-subject electroencephalogram analysis".

\subsection{How many training instances?}
\label{supplement: cvnumber}

To produce the results presented in this study, a total of more than {\num{100000}} different models were trained.
This value includes not only the models trained on the five cross-validation settings selected for this study, but also additional experiments that were performed to prepare some sections of the Supplementary Materials.
Furthermore, more than 80\% of the training instances originate from the full N-LOSO CV discussed in Section 3.3 of the paper.
This is also the reason why a full N-LOSO strategy is recommended only for datasets with at most 20 subjects.
The total number of instances for each cross-validation method and classification task is reported below:
\begin{align}
    \text{K-Fold} &= N_F \nonumber \\
    \text{LNSO} &= N_F \nonumber \\
    \text{LOSO} &= N_S \nonumber \\
    \text{N-LNSO} &= N_O \cdot N_I \nonumber \\
    \text{N-LOSO} &= N_S \cdot (N_S -1 ) \nonumber \\
    \text{N-Kfold} &= N_O \cdot N_F \nonumber \\
    \text{N-NLNSO} &= N_R \cdot N_O  \cdot N_I \nonumber
\end{align}
with $N_F$ the number of folds (10), $N_S$ the number of subjects (see Table 1 in the main body), $N_O$ the number of outer folds in nested cross-validations (10), $N_I$ the number of inner folds in nested cross-validations (10), and $N_R$ the number of times a cross-validation is repeated (10).
N-Kfold and N-NLNSO refer to the additional cross-validation settings explained in section A.4, which were used to provide a preliminary analysis of the generalization error underestimation with
traditional cross-validation methods.
Performance metrics for each trained models are shared within the openly available GitHub repository associated with this study.

\renewcommand{\arraystretch}{1.2}
\begin{table}
    \centering
    \caption{Balanced Accuracy median and $[25^{th}$$-75^{th}]$ percentiles for different random seeds.}
    \begin{tabular}{cccc}
        \toprule
        Task & Seed & LNSO & N-LNSO \\
        \bottomrule
        \multicolumn{4}{c}{ShallowConvNet} \\
        \toprule
        \parbox[t]{2mm}{\multirow{5}{*}{\rotatebox[origin=t]{90}{Parkinson}}} 
        & 0 &  82.85 $[75.02 - 89.99]$ & 75.78 $[66.44 - 84.86]$ \\ 
        & 42 &  85.02 $[81.80 - 88.65]$ & 80.48 $[74.51 - 86.30]$ \\ 
        & 1234 &  82.24 $[75.79 - 87.02]$ & 75.89 $[70.73 - 81.81]$ \\ 
        & 3407 &  82.80 $[75.20 - 93.05]$ & 76.30 $[69.70 - 89.00]$ \\ 
        & 83136297 &  85.38 $[71.77 - 88.91]$ & 77.10 $[67.79 - 82.95]$ \\ 
        \midrule
        \parbox[t]{2mm}{\multirow{4}{*}{\rotatebox[origin=t]{90}{Alzheimer}}} 
        & 0 &  54.20 $[52.53 - 62.47]$ & 53.01 $[49.19 - 57.09]$ \\ 
        & 42 & 62.41 $[51.01 - 63.70]$ & 55.87 $[50.42 - 61.43]$ \\
        & 1234 &  56.38 $[49.53 - 65.83]$ & 54.07 $[46.04 - 60.95]$ \\  
        & 3407 & 59.41 $[56.66 - 66.65]$ & 55.12 $[44.80 - 61.89]$ \\ 
        & 83136297 & 58.09 $[50.38 - 65.80]$ & 57.12 $[48.73 - 62.86]$ \\ 
        \midrule
        \parbox[t]{2mm}{\multirow{5}{*}{\rotatebox[origin=t]{90}{BCI}}} 
        & 0 &  55.49 $[53.31 - 58.35]$ & 54.19 $[51.12 - 56.02]$ \\ 
        & 42 &  56.61 $[52.66 - 58.37]$ & 54.06 $[51.79 - 56.13]$ \\ 
        & 1234 &  54.49 $[52.27 - 57.75]$ & 53.91 $[51.67 - 56.87]$ \\ 
        & 3407 &  56.07 $[54.19 - 57.09]$ & 54.46 $[52.13 - 56.36]$ \\ 
        & 83136297 &  54.72 $[53.39 - 58.91]$ & 53.41 $[50.91 - 56.20]$ \\ 
        \bottomrule
        \multicolumn{4}{c}{DeepConvNet} \\
        \toprule
        \parbox[t]{2mm}{\multirow{5}{*}{\rotatebox[origin=t]{90}{Parkinson}}} 
        & 0 &  61.97 $[51.26 - 83.32]$ & 54.72 $[51.44 - 66.39]$ \\ 
        & 42 &  74.16 $[66.34 - 82.88]$ & 65.12 $[52.17 - 73.34]$ \\ 
        & 1234 &  70.81 $[54.05 - 80.09]$ & 61.15 $[51.36 - 72.40]$ \\ 
        & 3407 &  63.52 $[50.57 - 72.76]$ & 56.26 $[50.57 - 70.98]$ \\ 
        & 83136297 &  53.35 $[51.81 - 81.40]$ & 53.28 $[50.77 - 68.15]$ \\
        \midrule
        \parbox[t]{2mm}{\multirow{5}{*}{\rotatebox[origin=t]{90}{Alzheimer}}} 
        & 0 &  56.98 $[51.94 - 62.05]$ & 52.51 $[46.36 - 56.90]$ \\
        & 42 &  57.10 $[47.17 - 60.28]$ & 51.68 $[46.19 - 56.31]$ \\ 
        & 1234 &  53.15 $[43.54 - 61.53]$ & 53.23 $[43.26 - 58.16]$ \\ 
        & 3407 &  50.17 $[44.97 - 57.11]$ & 52.00 $[43.36 - 56.98]$ \\ 
        & 83136297 &  58.78 $[52.01 - 61.68]$ & 54.38 $[48.08 - 59.69]$ \\ 
        \midrule
        \parbox[t]{2mm}{\multirow{5}{*}{\rotatebox[origin=t]{90}{BCI}}} 
        & 0 &  53.59 $[51.07 - 58.61]$ & 53.79 $[51.66 - 56.01]$ \\ 
        & 42 &  54.34 $[50.53 - 55.99]$ & 53.09 $[50.53 - 55.44]$ \\ 
        & 1234 &  55.35 $[52.98 - 57.24]$ & 53.12 $[51.46 - 55.09]$ \\ 
        & 3407 & 54.86 $[52.88 - 57.04]$ & 54.19 $[51.56 - 55.87]$ \\ 
        & 83136297 & 56.25 $[54.27 - 59.03]$ & 54.18 $[51.90 - 56.31]$ \\ 

        \bottomrule
    \end{tabular}
    \label{tab:seed}
\end{table}
\renewcommand{\arraystretch}{1}

\subsection{Random seed influence}
\label{supplement: seed}

To produce the results of this study, a custom random seed (83136297) was randomly selected and fixed.
This approach minimizes randomness in the code and improves the reproducibility of results.
While different random seeds may affect the reported accuracy scores, it can be assumed that this factor cannot alter the conclusions drawn in this research, especially given the relevant differences between the cross-validation methods.
To validate this, the LNSO and N-LNSO training instances for the ShallowNet and DeepConvNet models were repeated using 4 additional random seeds: 0, 42, 1234, and 3407.
Using a different random seed primarily impacts the following steps:
\WarningsOff
\begin{enumerate}[\textbullet]
 \item Subject assignment in the LNSO and N-LNSO cross-validation procedures.
 \item Creation of the data loader sampler iterator for mini-batch construction.
 \item Initialization of model weights.
\end{enumerate}
\WarningsOn

N-LOSO cross-validation was not included in this additional analysis due to the computational impracticality of rerunning this setting multiple times.
Moreover, considering the high variability among subjects, it can be inferred that the subject assignment step has the most significant impact on model accuracy.
This step is deterministic in N-LOSO since it evaluates every possible partition that assigns one subject to the validation set and another to the test set.
Table {\ref{tab:seed}} illustrates that changing the custom random seed does not affect the conclusions of this study.
The median values remain within the range of variability observed across different seeds for each model and task analyzed.
Additionally, pairwise comparisons of the median, $25^{th}$ percentile, and $75^{th}$ percentile balanced accuracies indicate that the metrics are higher in the LNSO cross-validation compared to the N-LNSO.

\subsection{Summary results with other metrics}
\label{supplement: metric}

\renewcommand{\arraystretch}{1.2}
\begin{table*}
    \centering
    \caption{Median and $[25^{th} - 75^{th}]$ percentiles of F1-score and Cohen's Kappa for each model, task, and cross-validation setting.}
    \begin{tabular}{cccccccc}
    \toprule
    \multirow{2}{*}{Task} & \multirow{2}{*}{Model} & \multicolumn{2}{c}{K-Fold} & \multicolumn{2}{c}{LNSO} & \multicolumn{2}{c}{N-LNSO} \\ \cline{3-8} 
    &  & \multicolumn{1}{c}{F1-score} & $\kappa$ & F1-Score & $\kappa$ & F1-Score & $\kappa$ \\
    \midrule
    \parbox[t]{2mm}{\multirow{5}{*}{\rotatebox[origin=d]{90}{Parkinson}}} 
    & ShallowConvNet
        & \makecell{1.00\\$[1.00-1.00]$}
        & \makecell{1.00\\$[1.00-1.00]$}
        & \makecell{0.85\\$[0.75-0.89]$}
        & \makecell{0.70\\$[0.44-0.77]$}
        & \makecell{0.76\\$[0.70-0.83]$}
        & \makecell{0.51\\$[0.37-0.64]$} 
    \\
    & EEGNet
        & \makecell{0.99\\$[0.99-0.99]$}
        & \makecell{0.99\\$[0.98-0.99]$}
        & \makecell{0.79\\$[0.70-0.87]$} 
        & \makecell{0.54\\$[0.40-0.72]$}
        & \makecell{0.72\\$[0.67-0.80]$} 
        & \makecell{0.42\\$[0.31-0.58]$} 
    \\
    & DeepConvNet
        & \makecell{1.00\\$[1.00-1.00]$}
        & \makecell{1.00\\$[1.00-1.00]$}
        & \makecell{0.52\\$[0.29-0.81]$}
        & \makecell{0.05\\$[0.04-0.62]$}
        & \makecell{0.52\\$[0.36-0.67]$}
        & \makecell{0.07\\$[0.01-0.35]$} 
    \\
    & T-ResNet
        & \makecell{0.85\\$[0.83-0.86]$} 
        & \makecell{0.70\\$[0.66-0.71]$}
        & \makecell{0.76\\$[0.61-0.80]$} 
        & \makecell{0.49\\$[0.25-0.59]$}
        & \makecell{0.65\\$[0.58-0.73]$} 
        & \makecell{0.28\\$[0.17-0.44]$} 
    \\
    \midrule
    \parbox[t]{2mm}{\multirow{5}{*}{\rotatebox[origin=d]{90}{Alzheimer}}} 
    & ShallowConvNet
        & \makecell{1.00\\$[0.99-1.00]$}
        & \makecell{0.99\\$[0.99-1.00]$}
        & \makecell{0.59\\$[0.53-0.68]$}
        & \makecell{0.39\\$[0.31-0.55]$}
        & \makecell{0.57\\$[0.50-0.65]$} 
        & \makecell{0.38\\$[0.27-0.51]$} 
    \\
    & EEGNet
        & \makecell{0.98\\$[0.97-0.98]$}
        & \makecell{0.96\\$[0.95-0.98]$} 
        & \makecell{0.55\\$[0.47-0.67]$} 
        & \makecell{0.35\\$[0.23-0.57]$}
        & \makecell{0.50\\$[0.44-0.60]$} 
        & \makecell{0.27\\$[0.18-0.50]$} 
    \\
    & DeepConvNet
        & \makecell{1.00\\$[0.99-1.00]$}
        & \makecell{0.99\\$[0.99-1.00]$} 
        & \makecell{0.55\\$[0.51-0.63]$}
        & \makecell{0.39\\$[0.28-0.44]$}
        & \makecell{0.53\\$[0.44-0.61]$} 
        & \makecell{0.32\\$[0.21-0.42]$} 
    \\
    & T-ResNet 
        & \makecell{0.78\\$[0.77-0.81]$} 
        & \makecell{0.66\\$[0.65-0.71]$} 
        & \makecell{0.45\\$[0.38-0.56]$} 
        & \makecell{0.16\\$[0.08-0.34]$} 
        & \makecell{0.44\\$[0.36-0.50]$}
        & \makecell{0.15\\$[0.03-0.24]$} 
    \\
    \midrule
    \parbox[t]{2mm}{\multirow{5}{*}{\rotatebox[origin=d]{90}{BCI}}} 
    & ShallowConvNet
        & \makecell{0.64\\$[0.63-0.66]$} 
        & \makecell{0.52\\$[0.51-0.55]$}
        & \makecell{0.54\\$[0.53-0.59]$}
        & \makecell{0.40\\$[0.38-0.45]$}
        & \makecell{0.53\\$[0.51-0.56]$} 
        & \makecell{0.38\\$[0.35-0.42]$} 
    \\
    & EEGNet
        & \makecell{0.57\\$[0.56-0.59]$} 
        & \makecell{0.43\\$[0.41-0.45]$}
        & \makecell{0.55\\$[0.54-0.56]$}
        & \makecell{0.40\\$[0.39-0.42]$}
        & \makecell{0.54\\$[0.51-0.57]$}
        & \makecell{0.39\\$[0.35-0.43]$} 
    \\
    & DeepConvNet
        & \makecell{0.65\\$[0.64-0.66]$} 
        & \makecell{0.54\\$[0.52-0.55]$}
        & \makecell{0.56\\$[0.54-0.59]$} 
        & \makecell{0.42\\$[0.39-0.45]$}
        & \makecell{0.54\\$[0.52-0.56]$}
        & \makecell{0.39\\$[0.36-0.42]$} 
    \\
    & T-ResNet 
        & \makecell{0.48\\$[0.47-0.50]$} 
        & \makecell{0.31\\$[0.30-0.34]$}
        & \makecell{0.47\\$[0.43-0.49]$}
        & \makecell{0.30\\$[0.26-0.32]$}
        & \makecell{0.45\\$[0.43-0.47]$} 
        & \makecell{0.28\\$[0.25-0.30]$} 
    \\
    \bottomrule
    \end{tabular}
    \label{tab:f1_and_kappa}
\end{table*}
\renewcommand{\arraystretch}{1}

This section summarizes the results of the sample-based K-Fold, Leave-N-Subjects-Out (LNSO), and Nested-Leave-N-Subjects-Out (N-LNSO) cross-validations using further performance metrics.
In particular, Table {\ref{tab:f1_and_kappa}} reports median and $[25^{th} - 75^{th}]$ percentiles for the F1-score and Cohen's Kappa.
Median values decreased when transitioning from the sample-based K-Fold method to LNSO, and further from LNSO to N-LNSO. 
Additionally, the interquartile range increases when comparing sample-based approaches to subject-based ones.
These performance variations reinforce the conclusions of this study and highlight the importance of employing nested approaches for evaluating EEG deep learning models.
For further insights, additional performance metrics, including Precision, Sensitivity, and the Area Under the Receiver Operating Characteristic Curve (ROC AUC), can be found in the tabular (.csv) file uploaded in the publicly accessible GitHub repository associated with this study.

\subsection{Generalization error underestimation with traditional cross-validation methods}
\label{supplement: bias}

\begin{figure*}[!t]
    \centering
    \includegraphics[width=\textwidth]{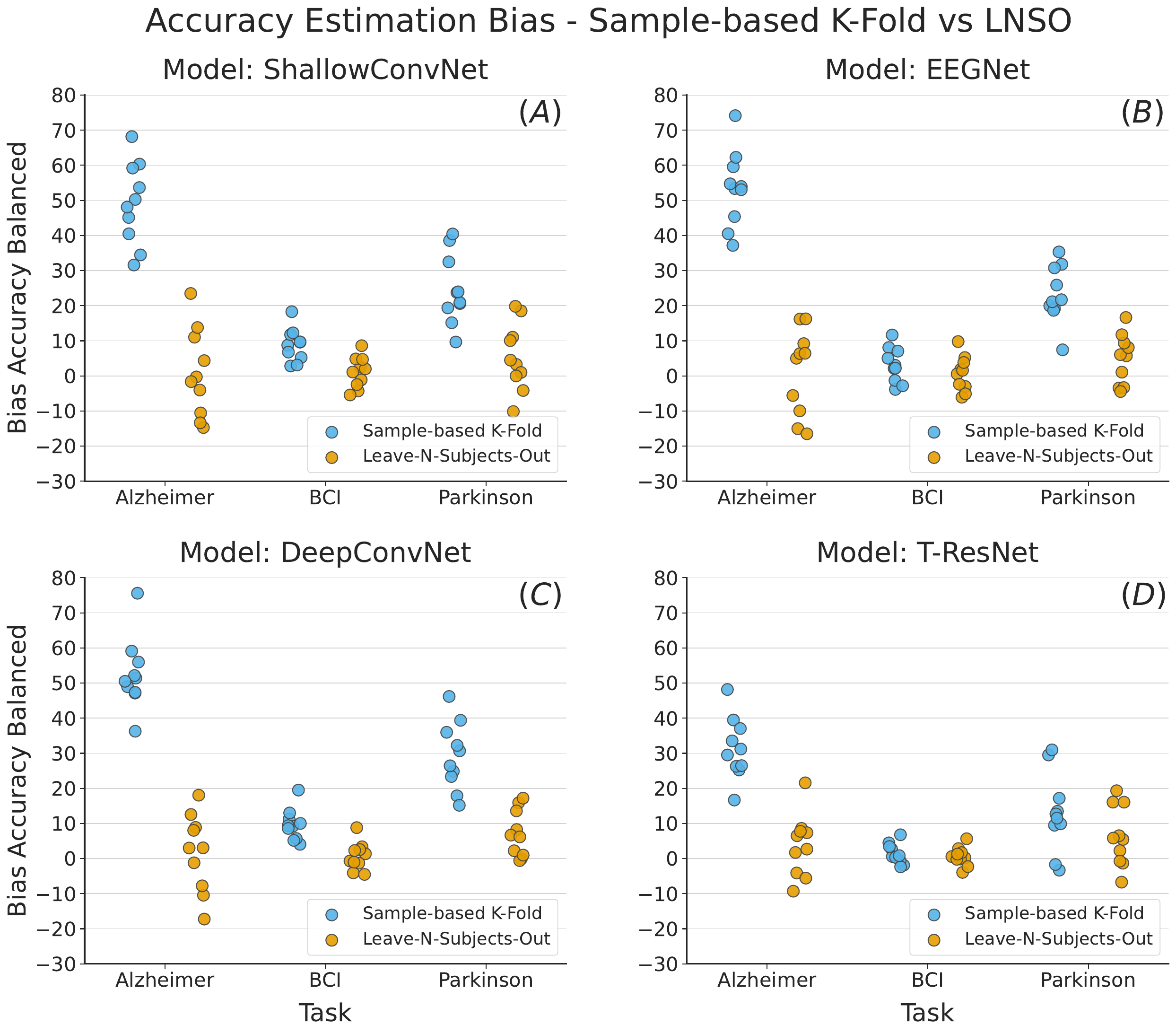}
    \caption{
    Balanced accuracy estimation bias comparison between Sample-based (K-Fold) and subject-based (LNSO) 10-Fold cross-validations.
    The sub-figures display results for different deep learning architectures: ShallowConvNet (Panel A), EEGNet (Panel B), DeepConvNet (Panel C), and T-ResNet (Panel D).
    Sample-based approaches produce strong positive biases, particularly in pathology classification tasks, indicating the tendency of this method to produce optimistic results and underestimating the generalization error.
    }
    \label{fig: biaskfoldvslnso}
\end{figure*}

To investigate whether significant differences among various cross-validation methods result from an overestimation of the generalization error by one approach or an underestimation by another, it is necessary to add an additional nesting level for each CV.
This involves repeating each cross-validation multiple times, with each run excluding a different group of subjects (10\% of the total) to use to measure the bias generalization error estimations.
During each run, the balanced accuracy is calculated for both the left-out group and the relevant CV evaluation split (i.e., the validation set for traditional CV and the test for nested approaches) for each trained model.
Subsequently, the mean difference is calculated.
In this preliminary analysis, ten repetitions of each CV have been performed, except for the LOSO vs N-LOSO scenario due to its substantial computational demands, as detailed in \autoref{supplement: cvnumber}.

In comparing the K-Fold with the LNSO, it is interesting to note that adding a nested level in traditional cross-validation methods is equivalent to run a nested cross-validation analysis, which supports the validity of the proposed method.
Furthermore, figure \ref{fig: biaskfoldvslnso} illustrates how sample-based methods tend to produce significant positive biases, resulting in underestimation of the generalization error.

In the N-LNSO versus LNSO comparison, the high variability and the limited number of bias estimations confirm the limitations of this preliminary approach.
Nevertheless, the mean and standard deviation of biases presented in Table \ref{tab:biaslnsonlsno} reveals to be generally slightly lower in the N-LNSO scenario.
Future analyses with more rigorously designed methodologies will offer further evidence and insights into the significant differences between the CV methods discussed in Section 3 and further elaborated in Section 4 of the paper.

\subsection{Additional Details on Model architectures}

This section offers detailed information about the deep learning models selected for this study.
A summary table is provided for each model, as indicated below: Table {\ref{tab:summaryEEGNet}} for EEGNet, Table {\ref{tab:summaryDeepConvNet}} for DeepConvNet, Table {\ref{tab:summaryShallow}} for ShallowConvNet, and Table {\ref{tab:summaryTResNer}} for T-ResNet.
Each table outlines important characteristics, including input shape, output shape, kernel size, number of convolutional groups, and the number of parameters for each layer in the network. 
Additionally, activation functions and dimensional operations (such as flattening and unsqueezing) are included for thoroughness.
The input and output sizes correspond to the Alzheimer's task.
Specifically, the models accept EEG trial windows consisting of 19 channels and 500 samples (equivalent to 4 seconds) and output probabilities for a three-class classification problem.
The variable N denotes the batch size, which was set to 64 for this study.
For clarity, skip connections in the Residual Blocks of T-ResNet are not illustrated.
The tables also facilitate a comparison of model complexity, with ShallowConvNet (Table {\ref{tab:summaryShallow}}) being the simplest model and T-ResNet (Table {\ref{tab:summaryTResNer}}) being the most complex.
A complete implementation of each deep learning model is accessible in the openly available SelfEEG library's code base (version 0.2.0).
Additionally, the GitHub repository associated with this study contains models exported in the Open Neural Network eXchange (ONNX) format.
These ONNX files can be imported into the Netron open platform ({\href{https://netron.app/}{https://netron.app/}}), allowing users to visualize how mini-batches pass through the layers of the network.

\renewcommand{\arraystretch}{1.2}
\begin{table}
    \centering
    \caption{Mean and standard deviation of the absolute balanced accuracy estimation bias for each task and model.}
    \begin{tabular}{cccc}
        \toprule
       Task & Model & LNSO & N-LNSO \\
        \midrule
        \parbox[t]{2mm}{\multirow{4}{*}{\rotatebox[origin=t]{90}{Parkinson}}} 
        & ShallowConvNet & $8.25\pm6.56$ & $8.11\pm5.35$\\
        & EEGNet & $6.99\pm4.38$ & $5.35\pm3.31$\\
        & DeepConvNet    & $7.17\pm6.15$ & $5.46\pm3.91$\\
        & T-ResNet       & $8.03\pm6.35$ & $6.54\pm3.91$\\
        \midrule
        \parbox[t]{2mm}{\multirow{4}{*}{\rotatebox[origin=t]{90}{Alzheimer}}} 
        & ShallowConvNet & $9.72\pm6.80$ & $9.26\pm6.46$\\
        & EEGNet & $10.64\pm4.60$ & $9.49\pm5.25$\\
        & DeepConvNet    & $9.02\pm5.47$ & $6.62\pm4.71$\\
        & T-ResNet       & $7.52\pm5.25$ & $5.42\pm3.86$\\
        \midrule
        \parbox[t]{2mm}{\multirow{4}{*}{\rotatebox[origin=t]{90}{BCI}}} 
        & ShallowConvNet & $3.71\pm2.21$ & $3.43\pm2.22$\\
        & EEGNet & $3.94\pm2.57$ & $3.17\pm2.49$\\
        & DeepConvNet & $3.00\pm2.30$ & $3.04\pm3.86$\\
        & T-ResNet & $1.88\pm1.74$ & $1.92\pm1.38$\\
        \bottomrule
    \end{tabular}
    \label{tab:biaslnsonlsno}
\end{table}
\renewcommand{\arraystretch}{1}

\texttt{
\begin{table*}
    \centering
    \caption{Summary Table: EEGNet. Input size reflects the Alzheimer's task.}
    \begin{tabular}{llllll}
        \toprule 
        Layer(type) & Input Shape & Output Shape & Kernel Shape & Groups & Param\# \\ 
        \midrule 
        EEGNet & $[\text{N},19,500]$ & $[\text{N},3]$ & -- & -- & -- \\ 
        +EEGNetEncoder & $[\text{N},19,500]$ & $[\text{N},240]$ & -- & -- & -- \\
        |\quad+Unsqueeze & $[\text{N},19,500]$ & $[\text{N},1,19,500]$ & -- & -- & -- \\ 
        |\quad+Conv2d & $[\text{N},1,19,500]$ & $[\text{N},8,19,500]$ & $[1,64]$ & 1 & 512 \\ 
        |\quad+BatchNorm2d & $[\text{N},8,19,500]$ & $[\text{N},8,19,500]$ & -- & -- & 16 \\ 
        |\quad+DepthwiseConv2d & $[\text{N},8,19,500]$ & $[\text{N},16,1,500]$ & $[19,1]$ & 8 & 304 \\ 
        |\quad+BatchNorm2d & $[\text{N},16,1,500]$ & $[\text{N},16,1,500]$ & -- & -- & 32 \\ 
        |\quad+ELU & $[\text{N},16,1,500]$ & $[\text{N},16,1,500]$ & -- & -- & -- \\ 
        |\quad+AvgPool2d & $[\text{N},16,1,500]$ & $[\text{N},16,1,125]$ & $[1,4]$ & -- & -- \\ 
        |\quad+Dropout & $[\text{N},16,1,125]$ & $[\text{N},16,1,125]$ & -- & -- & -- \\ 
        |\quad+SeparableConv2d & $[\text{N},16,1,125]$ & $[\text{N},16,1,125]$ & -- & -- & -- \\ 
        |\quad|\quad+DepthwiseConv2d & $[\text{N},16,1,125]$ & $[\text{N},16,1,125]$ & $[1,16]$ & 16 & 256 \\ 
        |\quad|\quad+ConstrainedConv2d & $[\text{N},16,1,125]$ & $[\text{N},16,1,125]$ & $[1,1]$ & 1 & 256 \\ 
        |\quad+BatchNorm2d & $[\text{N},16,1,125]$ & $[\text{N},16,1,125]$ & -- & -- & 32 \\ 
        |\quad+ELU & $[\text{N},16,1,125]$ & $[\text{N},16,1,125]$ & -- & -- & -- \\ 
        |\quad+AvgPool2d & $[\text{N},16,1,125]$ & $[\text{N},16,1,15]$ & $[1,8]$ & -- & -- \\ 
        |\quad+Dropout & $[\text{N},16,1,15]$ & $[\text{N},16,1,15]$ & -- & -- & -- \\ 
        |\quad+Flatten & $[\text{N},16,1,15]$ & $[\text{N},240]$ & -- & -- & -- \\ 
        +ConstrainedDense & $[\text{N},240]$ & $[\text{N},3]$ & -- & -- & 723 \\ 
        \bottomrule 
    \end{tabular}
    \label{tab:summaryEEGNet}
\end{table*}
}

\texttt{
\begin{table*}
    \centering
    \caption{Summary Table: DeepConvNet. Input size reflects the Alzheimer's task.}
    \begin{tabular}{llllll}
        \toprule 
        Layer(type) & InputShape & OutputShape & KernelShape & Groups & Param\# \\ 
        \midrule 
        DeepConvNet & $[\text{N},19,500]$ & $[\text{N},3]$ & -- & -- & -- \\ 
        +DeepConvNetEncoder & $[\text{N},19,500]$ & $[\text{N},200]$ & -- & -- & -- \\
        |\quad+Unsqueeze & $[\text{N},19,500]$ & $[\text{N},1,19,500]$ & -- & -- & -- \\ 
        |\quad+ConstrainedConv2d & $[\text{N},1,19,500]$ & $[\text{N},25,19,491]$ & $[1,10]$ & 1 & 275 \\ 
        |\quad+ConstrainedConv2d & $[\text{N},25,19,491]$ & $[\text{N},25,1,491]$ & $[19,1]$ & 1 & 11,900 \\ 
        |\quad+BatchNorm2d & $[\text{N},25,1,491]$ & $[\text{N},25,1,491]$ & -- & -- & 50 \\ 
        |\quad+ELU & $[\text{N},25,1,491]$ & $[\text{N},25,1,491]$ & -- & -- & -- \\ 
        |\quad+MaxPool2d & $[\text{N},25,1,491]$ & $[\text{N},25,1,163]$ & $[1,3]$ & -- & -- \\ 
        |\quad+Dropout & $[\text{N},25,1,163]$ & $[\text{N},25,1,163]$ & -- & -- & -- \\ 
        |\quad+ConstrainedConv2d & $[\text{N},25,1,163]$ & $[\text{N},50,1,154]$ & $[1,10]$ & 1 & 12,550 \\ 
        |\quad+BatchNorm2d & $[\text{N},50,1,154]$ & $[\text{N},50,1,154]$ & -- & -- & 100 \\ 
        |\quad+ELU & $[\text{N},50,1,154]$ & $[\text{N},50,1,154]$ & -- & -- & -- \\ 
        |\quad+MaxPool2d & $[\text{N},50,1,154]$ & $[\text{N},50,1,51]$ & $[1,3]$ & -- & -- \\ 
        |\quad+Dropout & $[\text{N},50,1,51]$ & $[\text{N},50,1,51]$ & -- & -- & -- \\ 
        |\quad+ConstrainedConv2d & $[\text{N},50,1,51]$ & $[\text{N},100,1,42]$ & $[1,10]$ & 1 & 50,100 \\ 
        |\quad+BatchNorm2d & $[\text{N},100,1,42]$ & $[\text{N},100,1,42]$ & -- & -- & 200 \\ 
        |\quad+ELU & $[\text{N},100,1,42]$ & $[\text{N},100,1,42]$ & -- & -- & -- \\ 
        |\quad+MaxPool2d & $[\text{N},100,1,42]$ & $[\text{N},100,1,14]$ & $[1,3]$ & -- & -- \\ 
        |\quad+Dropout & $[\text{N},100,1,14]$ & $[\text{N},100,1,14]$ & -- & -- & -- \\ 
        |\quad+ConstrainedConv2d & $[\text{N},100,1,14]$ & $[\text{N},200,1,5]$ & $[1,10]$ & 1 & 200,200 \\ 
        |\quad+BatchNorm2d & $[\text{N},200,1,5]$ & $[\text{N},200,1,5]$ & -- & -- & 400 \\ 
        |\quad+ELU & $[\text{N},200,1,5]$ & $[\text{N},200,1,5]$ & -- & -- & -- \\ 
        |\quad+MaxPool2d & $[\text{N},200,1,5]$ & $[\text{N},200,1,1]$ & $[1,3]$ & -- & -- \\ 
        |\quad+Dropout & $[\text{N},200,1,1]$ & $[\text{N},200,1,1]$ & -- & -- & -- \\ 
        |\quad+Flatten & $[\text{N},200,1,1]$ & $[\text{N},200]$ & -- & -- & -- \\ 
        +ConstrainedDense & $[\text{N},200]$ & $[\text{N},3]$ & -- & -- & 603 \\ 
        \bottomrule 
    \end{tabular}
    \label{tab:summaryDeepConvNet}
\end{table*}
}

\texttt{
\begin{table*}
    \centering
    \caption{Summary Table: ShallowConvNet. Input size reflects the Alzheimer's task.}
    \begin{tabular}{llllll}
        \toprule 
        Layer(type) & Input Shape & Output Shape & Kernel Shape & Groups & Param\# \\ 
        \midrule 
        ShallowNet & $[\text{N},19,500]$ & $[\text{N},3]$ & -- & -- & -- \\ 
        +ShallowNetEncoder & $[\text{N},19,500]$ & $[\text{N},1080]$ & -- & -- & -- \\
        |\quad+Unsqueeze & $[\text{N},19,500]$ & $[\text{N},1,19,500]$ & -- & -- & -- \\ 
        |\quad+Conv2d & $[\text{N},1,19,500]$ & $[\text{N},40,19,476]$ & $[1,25]$ & 1 & 1,040 \\ 
        |\quad+Conv2d & $[\text{N},40,19,476]$ & $[\text{N},40,1,476]$ & $[19,1]$ & 1 & 30,440 \\ 
        |\quad+BatchNorm2d & $[\text{N},40,1,476]$ & $[\text{N},40,1,476]$ & -- & -- & 80 \\
        |\quad+Square & $[\text{N},40,1,476]$ & $[\text{N},40,1,476]$ & -- & -- & 80 \\ 
        |\quad+AvgPool2d & $[\text{N},40,1,476]$ & $[\text{N},40,1,27]$ & $[1,75]$ & -- & -- \\ 
        |\quad+Log & $[\text{N},40,1,27]$ & $[\text{N},40,1,27]$ & -- & -- & -- \\
        |\quad+Dropout & $[\text{N},40,1,27]$ & $[\text{N},40,1,27]$ & -- & -- & -- \\ 
        |\quad+Flatten & $[\text{N},40,1,27]$ & $[\text{N},1080]$ & -- & -- & -- \\ 
        +Linear & $[\text{N},1080]$ & $[\text{N},3]$ & -- & -- & 3,243 \\ 
        \bottomrule 
    \end{tabular}
    \label{tab:summaryShallow}
\end{table*}
}

\texttt{
\begin{table*}
    \centering
    \caption{Summary Table: T-ResNet. Input size reflects the Alzheimer's task. Resnet blocks include a skip connection. }
    \begin{tabular}{llllll}
        \toprule 
        Layer(type) & Input Shape & Output Shape & Kernel Shape & Groups & Param\# \\ 
        \midrule 
        T-ResNet & $[\text{N},19,500]$ & $[\text{N},3]$ & -- & -- & -- \\ 
        +T-ResNetEncoder & $[\text{N},19,500]$ & $[\text{N},304]$ & -- & -- & -- \\
        |\quad+Unsqueeze & $[\text{N},19,500]$ & $[\text{N},1,19,500]$ & -- & -- & -- \\ 
        |\quad+Sequential & $[\text{N},1,19,500]$ & $[\text{N},16,19,250]$ & -- & -- & -- \\ 
        |\quad|\quad+Conv2d & $[\text{N},1,19,500]$ & $[\text{N},16,19,250]$ & $[1,7]$ & 1 & 112 \\ 
        |\quad|\quad+BatchNorm2d & $[\text{N},16,19,250]$ & $[\text{N},16,19,250]$ & -- & -- & 32 \\ 
        |\quad|\quad+ReLU & $[\text{N},16,19,250]$ & $[\text{N},16,19,250]$ & -- & -- & -- \\ 
        |\quad+Sequential & $[\text{N},16,19,250]$ & $[\text{N},16,19,250]$ & -- & -- & -- \\ 
        |\quad|\quad+ResBlock & $[\text{N},16,19,250]$ & $[\text{N},16,19,250]$ & -- & -- & -- \\
        |\quad|\quad|\quad+Conv2d & $[\text{N},16,19,250]$ & $[\text{N},16,19,250]$ & $[1,7]$ & 1 & 1,792 \\ 
        |\quad|\quad|\quad+BatchNorm2d & $[\text{N},16,19,250]$ & $[\text{N},16,19,250]$ & -- & -- & 32 \\ 
        |\quad|\quad|\quad+ReLU & $[\text{N},16,19,250]$ & $[\text{N},16,19,250]$ & -- & -- & -- \\ 
        |\quad|\quad|\quad+Conv2d & $[\text{N},16,19,250]$ & $[\text{N},16,19,250]$ & $[1,7]$ & 1 & 1,792 \\ 
        |\quad|\quad|\quad+BatchNorm2d & $[\text{N},16,19,250]$ & $[\text{N},16,19,250]$ & -- & -- & 32 \\ 
        |\quad|\quad|\quad+ReLU & $[\text{N},16,19,250]$ & $[\text{N},16,19,250]$ & -- & -- & -- \\ 
        |\quad|\quad+ResBlock & $[\text{N},16,19,250]$ & $[\text{N},16,19,250]$ & -- & -- & 3,648 \\ 
        |\quad|\quad+ResBlock & $[\text{N},16,19,250]$ & $[\text{N},16,19,250]$ & -- & -- & 3,648 \\ 
        |\quad+Sequential & $[\text{N},16,19,250]$ & $[\text{N},32,19,125]$ & -- & -- & -- \\ 
        |\quad|\quad+ResBlock & $[\text{N},16,19,250]$ & $[\text{N},32,19,125]$ & -- & -- & 14,528 \\ 
        |\quad|\quad+ResBlock & $[\text{N},32,19,125]$ & $[\text{N},32,19,125]$ & -- & -- & 14,464 \\ 
        |\quad|\quad+ResBlock & $[\text{N},32,19,125]$ & $[\text{N},32,19,125]$ & -- & -- & 14,464 \\ 
        |\quad|\quad+ResBlock & $[\text{N},32,19,125]$ & $[\text{N},32,19,125]$ & -- & -- & 14,464 \\ 
        |\quad+Sequential & $[\text{N},32,19,125]$ & $[\text{N},64,19,63]$ & -- & -- & -- \\ 
        |\quad|\quad+ResBlock & $[\text{N},32,19,125]$ & $[\text{N},64,19,63]$ & -- & -- & 57,728 \\ 
        |\quad|\quad+ResBlock & $[\text{N},64,19,63]$ & $[\text{N},64,19,63]$ & -- & -- & 57,600 \\ 
        |\quad|\quad+ResBlock & $[\text{N},64,19,63]$ & $[\text{N},64,19,63]$ & -- & -- & 57,600 \\ 
        |\quad|\quad+ResBlock & $[\text{N},64,19,63]$ & $[\text{N},64,19,63]$ & -- & -- & 57,600 \\ 
        |\quad|\quad+ResBlock & $[\text{N},64,19,63]$ & $[\text{N},64,19,63]$ & -- & -- & 57,600 \\ 
        |\quad|\quad+ResBlock & $[\text{N},64,19,63]$ & $[\text{N},64,19,63]$ & -- & -- & 57,600 \\ 
        |\quad+Sequential & $[\text{N},64,19,63]$ & $[\text{N},128,19,32]$ & -- & -- & -- \\ 
        |\quad|\quad+ResBlock & $[\text{N},64,19,63]$ & $[\text{N},128,19,32]$ & -- & -- & 230,144 \\ 
        |\quad|\quad+ResBlock & $[\text{N},128,19,32]$ & $[\text{N},128,19,32]$ & -- & -- & 229,888 \\ 
        |\quad|\quad+ResBlock & $[\text{N},128,19,32]$ & $[\text{N},128,19,32]$ & -- & -- & 229,888 \\ 
        |\quad+Sequential & $[\text{N},128,19,32]$ & $[\text{N},16,19,1]$ & -- & -- & -- \\ 
        |\quad|\quad+Conv2d & $[\text{N},128,19,32]$ & $[\text{N},16,19,26]$ & $[1,7]$ & 1 & 14,336 \\ 
        |\quad|\quad+AdaptiveAvgPool2d & $[\text{N},16,19,26]$ & $[\text{N},16,19,1]$ & -- & -- & -- \\ 
        +Linear & $[\text{N},304]$ & $[\text{N},3]$ & -- & -- & 912 \\ 
        \bottomrule 
    \end{tabular}
    \label{tab:summaryTResNer}
\end{table*}
}

\clearpage
\subsection{Nested cross-validation usage: differences between Machine Learning and Deep Learning}
\label{supplement: ML vs DL}

The analysis provided in this study supports the use of subject-based nested cross-validation methods to obtain more reliable performance estimates for EEG deep learning models.
Although this study primarily focuses on deep learning application, the validity of this approach can also be expanded to classical machine learning algorithms. 
However, the differences between the training paradigms betweem classical machine learning models and deep learning models lead to distinct motivations for using nested methods.

In classical machine learning, the validation set is not directly used during training for early stopping purposes.
Instead, the validation set plays a crucial role in the model’s hyperparameter tuning.
When a representative test set cannot be created (like in EEG data analysis), the recommended approach for evaluating model performance through cross-validation is to use nested approaches.
The nested level should be used for the hyperparameter tuning, while the outer level should be used to train a new model with the optimal hyperparameters using the entire training data. 
Additionally, classical machine learning models rely on hand-crafted features and typically have a lower number of parameters compared to the training set size, which means they are not overparameterized.
While overfitting can still occur, especially if sample-based partition methods are applied, it does not impact the quality of the learned features, as these are hand-crafted.
Furthermore, the selection of the validation set has minimal influence on the final cross-validation accuracy estimate, since regularization does not depend on monitoring validation error during training and the hyperparameters tuning considers the whole set of nested training instances.
However, the way folds are created (i.e., train/test splits) can still affect the final accuracy due to significant inter-subject variability.

In contrast, deep learning models used for EEG data analysis actively uses the validation set during training to restore the best weights when updates no longer yield an improvement in the validation loss (early stopping).
Unlike classical machine learning, deep learning does not use the nested level for hyperparameter tuning; rather, it assesses how the choice of validation set impacts test accuracy.
This is crucial in deep learning because the model automatically extracts features necessary for the task at hand through its encoder block.
Deep learning models are often overparameterized, which increases the risk of overfitting.
Combined with high inter-subject variability, this can significantly affect the quality of the learned features.
In this context, early stopping serves as a key regularization technique.
It is essential to have a validation set during training, and the effects of its selection must be carefully evaluated through nested approaches. 
Therefore, accuracies from the entire ensemble of trained models should contribute the final evaluation process.

In summary, while the concept of using a nested method is valid in both machine learning and deep learning, the role of the validation set in the nested level differs significantly between the two.
In classical machine learning, validation sets are not used for early stopping but are instead used for hyperparameter tuning.
Only the final model, retrained using the optimal hyperparameters on the entire training set (without a validation set), contributes to the final cross-validation performance estimate.
In deep learning, validation sets are integral for regularization through early stopping, and their creation can heavily influence test set accuracy.
Thus, accuracies from the complete ensemble of trained models are vital for determining final cross-validation performance estimates.

\subsection{Preliminary analysis with additional tasks}
\label{supplement: more tasks}

\renewcommand{\arraystretch}{1.2}
\begin{table*}[!h]
    \centering
    \caption{Median and $[25^{th} - 75^{th}]$ percentiles of Balanced accuracy and F1-score for the additional tasks (Sleep Deprivation and SSVEP). Results are reported for each model and cross-validation setting investigated.}
    \begin{tabular}{cccccccc}
    \toprule
    \multirow{2}{*}{Task} & \multirow{2}{*}{Model} & \multicolumn{2}{c}{K-Fold} & \multicolumn{2}{c}{LNSO} & \multicolumn{2}{c}{N-LNSO} \\ \cline{3-8} & &
    \multicolumn{1}{c}{Bal. Acc.} & \multicolumn{1}{c}{F1-score} & \multicolumn{1}{c}{Bal. Acc.} & \multicolumn{1}{c}{F1-score} & \multicolumn{1}{c}{Bal. Acc.} & \multicolumn{1}{c}{F1-score}
    \\
    \midrule 
    \parbox[t]{2mm}{\multirow{5}{*}{\rotatebox[origin=d]{90}{Sleep Deprivation $ $$ $}}} 
     & ShallowConvNet
        & \makecell{99.90\\$[99.84-100.00]$}
        & \makecell{1.00\\$[1.00-1.00]$}
        & \makecell{68.63\\$[58.80-74.47]$}
        & \makecell{0.69\\$[0.59-0.74]$}
        & \makecell{65.68\\$[58.73-73.08]$}
        & \makecell{0.65\\$[0.59-0.72]$}
    \\ 
     & EEGNet
        & \makecell{99.95\\$[99.90-100.00]$}
        & \makecell{1.00\\$[1.00-1.00]$}
        & \makecell{62.94\\$[53.86-65.97]$}
        & \makecell{0.59\\$[0.53-0.66]$}
        & \makecell{59.71\\$[52.50-65.16]$}
        & \makecell{0.58\\$[0.51-0.64]$}
    \\ 
     & DeepConvNet
        & \makecell{100.00\\$[99.92-100.00]$}
        & \makecell{1.00\\$[1.00-1.00]$}
        & \makecell{67.36\\$[56.74-69.59]$}
        & \makecell{0.67\\$[0.53-0.70]$}
        & \makecell{60.18\\$[55.19-66.03]$}
        & \makecell{0.57\\$[0.48-0.66]$}
    \\ 
     & T-ResNet
        & \makecell{75.77\\$[75.17-77.33]$}
        & \makecell{0.76\\$[0.75-0.77]$}
        & \makecell{63.06\\$[59.39-66.10]$}
        & \makecell{0.63\\$[0.59-0.66]$}
        & \makecell{59.57\\$[56.32-62.71]$}
        & \makecell{0.59\\$[0.55-0.63]$}
    \\ 
    \midrule 
    \parbox[t]{2mm}{\multirow{5}{*}{\rotatebox[origin=d]{90}{SSVEP}}} 
     & ShallowConvNet
        & \makecell{95.88\\$[95.66-96.15]$}
        & \makecell{0.96\\$[0.96-0.96]$}
        & \makecell{94.28\\$[92.99-95.84]$}
        & \makecell{0.94\\$[0.93-0.96]$}
        & \makecell{92.64\\$[89.90-94.60]$}
        & \makecell{0.93\\$[0.90-0.95]$}
    \\ 
     & EEGNet
        & \makecell{97.62\\$[97.25-97.89]$}
        & \makecell{0.98\\$[0.97-0.98]$}
        & \makecell{97.75\\$[96.59-98.44]$}
        & \makecell{0.98\\$[0.97-0.98]$}
        & \makecell{97.18\\$[95.34-98.05]$}
        & \makecell{0.97\\$[0.95-0.98]$}
    \\ 
     & DeepConvNet
        & \makecell{97.43\\$[97.00-97.88]$}
        & \makecell{0.97\\$[0.97-0.98]$}
        & \makecell{97.38\\$[96.36-98.22]$}
        & \makecell{0.97\\$[0.96-0.98]$}
        & \makecell{96.52\\$[94.07-97.47]$}
        & \makecell{0.97\\$[0.94-0.97]$}
    \\ 
     & T-ResNet
        & \makecell{95.49\\$[95.23-95.59]$}
        & \makecell{0.95\\$[0.95-0.96]$}
        & \makecell{97.42\\$[93.69-97.79]$}
        & \makecell{0.97\\$[0.94-0.98]$}
        & \makecell{96.45\\$[93.05-97.25]$}
        & \makecell{0.96\\$[0.93-0.97]$}
    \\ 
    \bottomrule 
    \end{tabular}
    \label{tab:more_task}
\end{table*}
\renewcommand{\arraystretch}{1}

\renewcommand{\arraystretch}{1.2} 
\begin{table}[!h]
\centering
\caption{Learning rate grid for the additional tasks.}
\begin{tabular}{cccc}
\toprule
Model & \makecell{Sleep\\Deprivation} & SSVEP \\
\midrule
ShallowConvNet 
    & $5.0\cdot10^{-5}$
    & $7.5\cdot10^{-4}$
\\
EEGNet
    & $1.0\cdot10^{-3}$
    & $7.5\cdot10^{-4}$
\\
DeepConvNet
    & $2.5\cdot10^{-4}$
    & $1.0\cdot10^{-3}$
\\
T-ResNet
    & $5.0\cdot10^{-5}$
    & $1.0\cdot10^{-3}$
\\
\bottomrule
\end{tabular}
\label{tab: learningrate2}
\end{table}
\renewcommand{\arraystretch}{1}

This section provides a preliminary analysis for the LNSO vs. K-Fold (Subsection 3.1) and the LNSO vs. N-LNSO (Subsection 3.1) comparisons with two additional tasks, namely:

\WarningsOff
\begin{enumerate}[\textbullet]
    \item \textit{Sleep deprivation}: a binary classification task. It aims to distinguish between subjects in normal or sleep deprived states. Sleep deprived EEGs are widely used for investigation of patients who have seizures or blackouts, representing an important tool for the diagnosis of this neurological disease.
    \item \textit{Steady State Visually Evoked Potential (SSVEP)}: a four-class classification task. It aims to discriminate between responses to different stimuli at specific frequencies (5.45, 6.67, 8.57, and 12 Hz) that were presented in four positions (down, right, left, and up, respectively). SSVEP EEGs are used in a variety of settings, including brain-computer interface applications (e.g., wheelchair control systems) and clinical applications in neuroscience.
\end{enumerate}
\WarningsOn

The following sections briefly describe the data sets used to construct these tasks.

\subsubsection{ds004902 - Sleep Deprivation}
This dataset\footnote{
    D. Salisbury, D. Seebold, and B. Coffman, “EEG: First episode psychosis vs. control resting task 2,” 2022. \href{https://doi.org/10.18112/openneuro.ds003947.v1.0.1}{doi:10.18112/openneuro.ds003947.v1.0.1}.
}, selected from the OpenNeuro platform, contains resting-state eyes open/closed EEG recordings from 71 healthy subjects (age $20.0 \pm 1.4$ years).
All the subjects underwent two recording sessions (duration $4.9 \pm 0.4$ minutes).
One in a sleep deprived state and the other in normal sleep state. 
Not all subjects have eyes closed recordings, so only eyes open recordings were used.
Raw data were acquired with a 61 channel system (10/10 template, reference on FCZ) at 500 Hz.
Preprocessing was performed with the same steps as for the Parkinson's datasets (see Table 2 in section 2.2).
The total number of samples is {\num{9914}}.

\subsubsection{SSVEP}
This dataset\footnote{
    M.-H. Lee, O.-Y. Kwon, Y.-J. Kim, H.-K. Kim, Y.-E. Lee, J. Williamson, S. Fazli, S.-W. Lee, EEG dataset and OpenBMI toolbox for three BCI paradigms: an investigation into BCI illiteracy, GigaScience 8 (5) (2019) giz002. \href{https://doi.org/10.1093/gigascience/giz002}{doi:10.1093/gigascience/giz002}.
}, selected from the GigaDB platform, contains steady-state visually evoked potential EEG data from 54 healthy subjects (age $24.2 \pm 3.0$ years).
All the subjects underwent two recording sessions divided in two separate phases.
Each phase contained 100 trials, 25 per class, associated with the response to different stimuli at specific frequencies (5.45, 6.67, 8.57, and 12 Hz) that were presented in four positions (down, right, left, and up, respectively).
Each trial lasted 4 seconds.
Raw data were acquired with a 62 channel system (10/10 Template, nasion-referenced and grounded to electrode AFz) at {\num{1000}} Hz.
Preprocessing was performed with the same steps as for the BCI datasets (see Table 2 in section 2.2).
The total number of samples is {\num{21600}}.

Learning rates were selected using the same procedure described in subsection 2.4.1 and are listed in Table {\ref{tab: learningrate2}}.
Other values of the training hyperparameters (e.g., seed, batch size, optimizer) were left unchanged.

Table {\ref{tab:more_task}} reports accuracy and F1-score metrics. Sleep deprivation results further support the conclusions of this study.
The median values decrease when moving from the sample-based K-fold method to LNSO, and further from LNSO to N-LNSO. 
In addition, the interquartile range increases when sample-based approaches are compared to subject-based approaches.
The SSVEP task appears to produce very high cross-subject accuracy and F1 scores, making it difficult to compare cross-validation methods.
Nevertheless, this preliminary analysis reinforces the conclusions of this study and highlights the importance of using nested approaches to evaluate EEG deep learning models.



\end{document}